\documentclass[a4paper,11pt]{article}
\usepackage{jheppub}
\usepackage{graphicx}
\usepackage{amsmath}
\usepackage{amsfonts}
\usepackage{amssymb}
\usepackage{paralist}
\usepackage[colorlinks=true,linkcolor=blue]{hyperref}

\def\half{\frac 12}
\def\siml{{\ \lower-1.2pt\vbox{\hbox{\rlap{$<$}\lower6pt\vbox{\hbox{$\sim$}}}}\ }}
\def\simg{{\ \lower-1.2pt\vbox{\hbox{\rlap{$>$}\lower6pt\vbox{\hbox{$\sim$}}}}\ }}

\def\putbox#1#2{\includegraphics[width=#1\textwidth]{#2}}

\def\Eq#1{Eq.~(\ref{#1})}
\def\Fig#1{Fig.~\ref{#1}}
\def\be{\begin{equation}}
\def\ee{\end{equation}}
\def\bea{\begin{eqnarray}}
\def\eea{\end{eqnarray}}
\def\ncond{n_{\mathrm{cond}}}
\def\kcond{k_{\mathrm{cond}}}
\def\kinit{k_{\mathrm{init}}}
\def\mth{\omega_{\mathrm{th}}}
\def\mtha{\omega_{\mathrm{th},a}}
\def\OO{\mathcal{O}}
\def\ZZ{\mathcal{Z}}
\def\RR{\mathcal{R}}
\def\Re{\,\mathrm{Re}\,}

\title{Condensates in Relativistic Scalar Theories}

\author{Guy D.\ Moore}

\affiliation{Institut f\"ur Kernphysik, Technische Universit\"at Darmstadt\\
  Schlossgartenstra\ss e 2, D-64289 Darmstadt, Germany}

\abstract{Scalar field theory with large infrared initial
  occupancy develops very large deep-infrared occupancy, which locally
  resembles a Bose-Einstein condensate.  We study the structure and
  spatial coherence of this condensate.  The $O(N)$ symmetric theory
  with $N>1$ is qualitatively different than $N=1$.  We explain the
  thermodynamical reason why, for $N>1$, the
  condensate locally carries nearly maximal conserved charge density.
  We also show how this property impedes the condensate's decay, and
  we show that it prevents the condensate from ever becoming fully
  spatially homogeneous.  For $N\leq 4$ the condensate can carry
  topological defects, but these do not appear to control the
  large-$k$ tail in its power spectrum, which is the same for $N=8$
  where there are no topological defects.}


\begin{document}
\maketitle

\section{Introduction}
\label{sec:intro}

Relativistic scalar fields with strongly infrared (IR) initial
conditions have been studied for their possible role in post-inflation
cosmology \cite{%
  Kofman:1994rk,%
  Khlebnikov:1996mc,%
  Prokopec:1996rr,%
  Micha:2002ey,%
  Micha:2004bv%
}
and as an analogue theory to QCD
to study the problem of thermalization immediately after a heavy ion
collision.  There are physical differences between IR-occupied scalars
and nonabelian gauge fields, most notably the role of much more
efficient particle-number changing processes in the gauge theory
\cite{AMY5}, which prevent the formation of infrared condensates
\cite{Kurkela:2012hp}.  Nevertheless, the theory is of considerable
interest for its own merits.

Particularly interesting is the rich infrared physics which is present
in a scalar theory when the initial state features large occupancy of
relatively IR modes.  This problem has been studied
extensively for scalar fields with $N$ components,
\cite{%
  Berges:2008wm,
  Berges:2010ez,
  Gasenzer:2011by,
  Berges:2012us,
  Schmidt:2012kw,
  Berges:2012ev,
  Nowak:2013juc,
  Karl:2013kua,
  berg2,
  Orioli:2015dxa
}
using the tools of classical (statistical) field theory, which should
be a good approximation in a regime where the occupancy is large.
They have found several scaling behaviors in different wave number
ranges, some of which have clear explanations in terms of kinetic or
2-particle irreducible descriptions \cite{Berges:2010ez}.  Here we will
study in a little more detail the behavior of the most infrared modes, well
below the effective thermal oscillation frequency induced by mode-mode
interactions.  Berges \textsl{et al} demonstrated three scaling
regimes \cite{Berges:2010ez,Orioli:2015dxa}.  Below a scale
$\kcond \propto t^{-1/2}$ the occupancy is very large and at most
weakly $k$-dependent%
\footnote{Ref.\ \cite{Orioli:2015dxa} find an $f(k)\propto k^{-1/2}$
  dependence in the deep infrared.  We find $k^0$.  At this
  time we do not understand the origin of this discrepancy.};
then it falls
steeply, roughly as $k^{-5}$, until it softens to a $k^{-3/2}$
``energy-cascade'' scaling. (Here $k$ is the wave number%
\footnote{%
  Since we use a classical framework, we will talk about wave numbers
  and avoid conflating them with momenta.}
and $f$ is a particle occupancy estimate
based on the $k$ Fourier component of the field and its time
derivative.)  We will be interested in these two most-infrared
regions.  The large occupancy in the deep IR can be
understood as a consequence of an approximate particle number
conservation.  Thermodynamically, the excess particle number in the
initial conditions tries to organize into an IR condensate, but by
causality it cannot instantly fall into a single $k=0$-mode across a
large system.

In studying this IR occupancy, we will
concentrate on cases with ``large'' condensates, where the
particle number in the deep IR is comparable to the
particle number in all other modes, so the IR modes dominate the
effective thermal oscillation frequency.%
\footnote{%
  The energy is $\varepsilon \sim \int d^3 k\; \omega_k f(k)$;
  particle number is $n \sim \int d^3 k\; f(k)$, and contribution to
  dispersion corrections is $\sim \int d^3 k\; \omega^{-1} f(k)$.
  If an $\OO(1)$ fraction of particles are in the IR, they play a
  small role in energy but they dominate dispersion corrections.}
We will refer to the large occupancy in very IR wave numbers
as a ``local condensate,'' since by local probes it appears to be a
condensate and it occurs for the same (particle-number storage)
reasons as a true condensate, but it lacks the system-scale long-range
order of a true IR condensate, which should reside in one mode.
This local condensate then undergoes ordering dynamics, in which it
evolves towards higher long-range order.  The $k$-spectrum is a
Fourier representation of this ordering dynamics.

We will show that the dynamics of the local condensate depends
nontrivially on the number of field components $N$ (we only consider
multi-component fields which possess global $O(N)$ symmetry).  The
case $N=1$ is qualitatively different than all cases $N \geq 2$.  We
will explain why, for $N \geq 2$,
the condensate locally carries large -- nearly
maximal -- conserved charge density.  This phenomenon has previously
been observed in the $N=2$ case \cite{Gasenzer:2011by}.
The large local charge density causes the condensate to
decay much more slowly for $N>1$ than for the $N=1$ case.  It also
means that, in a finite box, the condensate eventually becomes spatially
homogeneous for $N=1$ but it remains spatially varying for $N\geq 2$,
with overall charge neutrality enforcing that the condensate vary in
such a way that its total charge is zero.  For several $N$ values the
ordering dynamics include networks of topological defects, including
domain walls for $N=2$ \cite{Gasenzer:2011by} and strings for
$N=1,2,3$ (as we will show).  These topological features do not
control the high-$k$ tail of the power spectrum, however; this tail is
the same when comparing theories with such defects to theories with no
defects such as $N=8$.

In the remaining sections we will explain the physics of the results
described above, and we will support the description with results from
3+1 dimensional classical (statistical) lattice field theory
simulations using $N=1$, 2, 4, and 8, with lattices of $512^3$ and
$128^3$ points (the former for spectra, the latter to study how the
condensate moves into the $k=0$ and $\vec k=\frac{2\pi}{L}[1,0,0]$
modes).

\section{General picture}
\label{general}

Consider $O(N)$ scalar field theory with Lagrangian
\be
\label{Lagrangian}
- \mathcal{L} =
\sum_a \left( \half \partial_\mu \phi_a \partial^\mu \phi_a \right)
+ \sum_a \frac{m^2}{2} \phi_a^2 + \frac{\lambda}{8}
  \left( \sum_a \phi_a^2 \right)^2 \,,
\ee
with initial conditions which put a large initial occupancy in IR
modes.  For suitably large occupancy the fields can be treated as
classical.  At the lowest order, each $k$-mode oscillates
independently at frequency $\omega_{k,a} = \sqrt{k^2 + \mtha^2}$,
where the effective thermal oscillation frequency of species $a$ is
$\mtha^2 \simeq m^2 + \frac{\lambda}{2}
\langle 2\phi_a^2+\sum_b \phi_b^2 \rangle$.
At next order, the modes interact with each other,
exchanging energy between modes and into formerly unoccupied modes.
Entropy considerations tell us that the energy will spread
out into all available modes, which for phase-space reasons are
dominated by larger-$k$ modes.  Scalar theory has an
approximately-conserved particle number; for a $k$-mode with
oscillation frequency $\omega_k = \sqrt{k^2+\mth^2}$ and carrying
energy $\varepsilon(k)$, the particle number is
$f(k) = \varepsilon(k)/\omega_k$.  This particle number is not
strictly conserved, but it decays on a much longer
time scale than the kinetic equilibration needed to re-arrange the
energy between $k$-modes.  Now $\omega_k$ in UV modes is larger
than in the modes occupied in the initial conditions; so as these
modes absorb the system's energy, they take little of its particle
number, leaving an excess in the IR modes.

Naively, we might expect the modes to take an occupancy of form
\be
\label{fakeEQ}
f(k) \sim \frac{T}{\omega_k - \mu}
\ee
with $T$ and $\mu$ the Lagrange multipliers to conserve the
system's energy and particle number.  For sufficiently large initial
$n/\varepsilon$ ratio, $\mu=\mth$ is not enough to contain all the
particle number in the finite-$k$ modes, and an $\OO(1)$ fraction of
the particle number can go into the mode or modes with the smallest
$\omega_k$ value, which is permitted by \Eq{fakeEQ} if
$\mu=\omega_0$.  In practice, \Eq{fakeEQ} can occur at late times if a
lattice or other physics provides a UV cutoff, but generally the
occupancies take a more complicated form.  But the ``need'' to store a
large particle number is a fairly general issue, and a large occupancy
in a small $k$-space region is a general solution.

For statistically uniform initial conditions over a large system,
this thermodynamic preference for large occupancy at small $k$ occurs
everywhere.  Therefore, locally in the system -- on scales large
compared to $\mth^{-1}$ but small compared to the system size -- particle
number should move into a nearly-uniform condensate.  But since this
occurs due to local physics in a time scale too short for information
to be exchanged throughout the system, this condensate will not
instantly form coherently -- in the same field direction and with the
same oscillation phase -- everywhere in the system at once.  Instead, it
will generically form with an independent field direction and
oscillation phase at widely spatially separated points.  This similar
to a quench process which leads to a vacuum expectation value (VEV), a
common phenomenon in condensed matter physics (see for instance
\cite{Mondello:1990zz,Bray}) which has also been studied in the context
of cosmological phase transitions
\cite{Kibble:1976sj,Vilenkin:1984ib,Turok:1989ai}.

A uniform true condensate would have a common oscillation phase and
field direction throughout the system.  This would spontaneously break
the symmetry between different oscillation phases and field
directions.  Therefore the formation of a condensate is a type of
symmetry-breaking transition.  This is another feature in common with
quench processes.  But, just as for the formation of a VEV, the
local condensate initially chooses its phase and direction
independently at widely spaced points.  The subsequent dynamics
involve the organization of the initially nonuniform features.  It is
well known that the details of this field organization depend on
the space of possible local values (field direction and oscillation
phase) for the condensate, which plays the role of the
vacuum manifold in the usual case.  To simplify the language, we will
call the space of possible local values for the condensate the
CCspace (condensate configuration space).  The topology of this space
determines what types of topological structures can occur and impede
the ordering dynamics.

Therefore our first order of business is to determine the possible
values a condensate can take locally, and to understand the topology
of the CCspace.

\section{Describing the Local Condensate}
\label{sec:local}

The space of field directions and phases which the condensate can take
depends in an important way on $N$ the number of field components.
For $N=1$ the theory has no continuous global symmetry, which makes
this case essentially different from $N \geq 2$.  Therefore we start
by considering $N=1$ and then turn to $N \geq 2$.

\subsection{N=1}
\label{secN1}

The condensate is described instantaneously by the field value and its
time derivative.  This is equivalent to the field's peak amplitude
$\phi_0$ and the phase of its oscillation,
$\phi = \phi_0 \Re e^{i(\varphi+\omega t)}$, where $\varphi$ is
defined modulo $2\pi$.  The amplitude is fixed by the particle
number density which must enter the condensate, so the variable which
can differ through space is the phase $\varphi$.  Therefore the
CCspace has the topology of the circle $S^1$.

We also compute the relation between the condensate's peak amplitude
$\phi_0$, particle number density $n$, energy density $\varepsilon$,
and oscillation frequency $\omega$, assuming that the condensate's
self-interactions dominate its interactions with other fluctuations in
establishing the oscillation frequency.  Saving the details for
Appendix \ref{appomega}, we find that
\bea
\label{omega1}
\omega &=& \frac{\sqrt{\pi}\, \Gamma(3/4)}{\Gamma(1/4)}
\sqrt{\lambda} \phi_0 \simeq 0.59907 \sqrt{\lambda} \phi_0
\\
\nonumber
& = & 1.00751 \lambda^{1/4} \varepsilon^{1/4} \,,
\\
\label{energy1}
\varepsilon & = & \frac{3 \sqrt{\pi} \Gamma(3/4)}{2^{5/4} \Gamma(1/4)}
\lambda^{1/3} n^{4/3} \simeq 0.68825 \lambda^{1/3} n^{4/3} \,.
\eea

\subsection{N greater than 1}
\label{secN2}

Next consider the case of a field with $N\geq 2$ components.  The
instantaneous value of the condensate is determined by the (locally
space-averaged) instantaneous value of $\phi_a$ and $\dot{\phi}_a$.
One possibility is that $\phi_a$ and $\dot{\phi}_a$ lie in the same
field direction.  In this case the field oscillates along one field
direction as in the $N=1$
case.  The other extreme is for $\dot\phi_a$ to be orthogonal in field
space to $\phi_a$, and of a magnitude which keeps $|\phi|$ unchanged
with time.  That is, the scalar field can follow a circular orbit
through $O(N)$ field space.

The condensate has to carry a certain particle number density.
Statistical mechanics arguments favor whichever form for the
condensate can do so with minimum energy.  Therefore it is important
to repeat the calculation of $\omega$, $\varepsilon$, and $n$ as a
function of $\phi_0$ for the circular orbit case.  As shown in
Appendix \ref{appomega}, for this case we have
\bea
\label{omega2}
\omega & = & \frac{\lambda^{1/2}}{2^{1/2}} \phi_0 = 
\frac{2^{1/4}}{3^{1/4}} \lambda^{1/4} \varepsilon^{1/4}
\simeq 0.9036 \lambda^{1/4} \varepsilon^{1/4} \,,
\\
\label{energy2}
\epsilon &=& \frac{3}{2^{7/3}} \lambda^{1/3} n^{4/3}
\simeq 0.5953 \lambda^{1/3} n^{4/3} \,.
\eea
The energy cost of a circular orbit is about $13\%$ lower, at fixed
particle number density, than back-and-forth oscillation.  Of course,
there are also possibilities intermediate between these, corresponding
to (precessing) elliptical orbits in field space.  These carry energy
per particle number strictly intermediate between the two limiting
cases we have considered, so the circular-orbit case is the most
energetically efficient way to store particle number.  Therefore it
will be favored on statistical-mechanical grounds.%
\footnote{Another way to see this is to consider a generic elliptical
  orbit and to consider the energy cost of adding a quantum in the
  direction which stretches the ellipse, versus adding a quantum in
  the direction which makes it more circular.  The energy cost is
  lower for the quantum which makes the oscillation more circular.}

The circular orbit carries a large local density of at least one of
the conserved charges
$\rho_{ab} = \dot\phi_a \phi_b - \dot\phi_b \phi_a$ ($a<b$).
One might expect this to forbid circular orbits, since it requires
large charge densities and charge is conserved.
But charge conservation is global, and the condensate can
exchange charge with fluctuations, which propagate freely into other
regions and deposit charge density with the condensate there.
Therefore the condensate can take on a large charge density
locally, provided that the charge density vary through space such that
its space integral vanishes.  In the next section we will present
numerical evidence showing that this is what occurs.

How do we describe the condensate locally?  It is described by the
field direction $\phi_a$ and the field-derivative direction
$\dot\phi_a$, with the constraints that each is of fixed
magnitude (to carry the correct particle number density and maintain
minimum energy cost) and that $\dot\phi_a$ is orthogonal to $\phi_a$.
A fixed-length $\phi_a$ is an element of the
$(N{-}1)$-sphere $S^{N-1}$, while a fixed-length orthogonal $\dot\phi_a$
is an element of the fixed-length tangent bundle.  Therefore the
CCspace is topologically the unit-tangent bundle of $S^{N-1}$,
$UT(S^{N-1})$, which is a fibration of $S^{N-2}$ over $S^{N-1}$.

\subsection{Topological considerations}
\label{sectop}

A nontrivial CCspace can have consequences for the ordering dynamics.
If the CCspace is not connected -- if it has a nontrivial $\pi_0$
homotopy group -- the condensate in different regions may be separated
by domain walls.  If it is not simply connected -- if the first
homotopy group $\pi_1$ is nontrivial -- it can vary in a topologically
nontrivial way around a loop in field space, guaranteeing the
existence of string defects.  If $\pi_2$ is nontrivial, there can be
monopole defects.  Let us see which occur for some values of $N$:
\begin{enumerate}
\item \label{caseN1}
  For $N=1$ the CCspace is topologically the circle $S^1$.  This has
  $\pi_1(S^1) = \ZZ$ the integers.   Therefore there are string
  defects, corresponding to lines where the condensate's phase changes
  by $2\pi$ as one goes around the line.
\item \label{caseN2}
  The case $N=2$ is equivalent to a complex scalar field.  The CCspace
  is the unit-tangent bundle of the circle $S^1$.  The unit-tangent
  space has two points, corresponding to the condensate rotating
  clockwise or counterclockwise around the complex plane.  So the
  unit-tangent bundle is two copies of the circle, $S^1 \times \ZZ_2$.
  Again $\pi_1 = \ZZ$ and there are strings; but also $\pi_0 = \ZZ_2$
  and there are domain walls, separating regions where the condensate
  revolves clockwise from regions where it revolves counterclockwise.
  This is the only $N$ value for which there are domain walls.
\item \label{caseN3}
  For $N=3$ the unit-tangent bundle $UT(S^2)$ is equivalent to $SO(3)$
  the group of 3-dimensional rotations.  To see why, note that
  $\phi_a$ is a direction in $\RR^3$, while $\dot\phi_a$ must be
  another $\RR^3$ direction orthogonal to the first.  Together with
  their cross product, they define an orthogonal coordinate frame.  The
  space of coordinate frames is the same as the space of rotations
  (think of the rotation from a standard frame to the desired frame),
  which is $SO(3)$.

  It is important that $SO(3) \neq SU(2)$ its double cover.  In
  particular, since it is double-covered, $\pi_1(SO(3)) = \ZZ_2$ and
  there are string defects.  However $\pi_0(SO(3))$ and $\pi_2(SO(3))$
  are trivial so there are no domain walls or monopoles.
\item \label{caseN4}
  For $N=4$, the tangent bundle of $S^3$ is trivial, so
  $UT(S^3) = S^2 \times S^3$.  Therefore $\pi_0$ and $\pi_1$ are
  trivial, but $\pi_2(UT(S^3)) = \ZZ$ and there are monopole defects.
\item \label{caseNlarge}
  For general $N$ the unit-tangent bundle is
  $UT(S^{N-1}) = \mathrm{Spin}(N)/\mathrm{Spin}(N-2)$.
  For $N>4$ this space has $\pi_0$, $\pi_1$, and $\pi_2$ trivial, so
  there are no walls, strings, or monopoles.  In some dimensions
  $\pi_3$ is nontrivial and there are textures, but for the special
  case $N=8$, $UT(S^7) = S^6 \times S^7$, which has no homotopy below
  $\pi_6$ and so is free of defects
  in 3+1 dimensional space.%
  \footnote{%
    I thank Johannes Walcher for a refresher on unit tangent bundles.}
\end{enumerate}

\section{Effects of Condensate Structure}
\label{seceffects}

Here we will look at the consequences of the form of the condensate,
and check them against numerical investigations.

\subsection{Local charge density}
\label{secdens}

The most convincing evidence that the above description of
condensation is true is to look at the process in a small enough
volume that it reaches completion.  Then we analyze the late-time
behavior and see that it is qualitatively different for the cases
$N=1$, $N=2$, and $N>2$.

For $N=1$ there is no obstacle to the condensate developing completely
in the $k=0$ mode.  But for $N>1$ the condensate locally carries a net
abundance of conserved charge, which should average to zero globally.
Therefore the condensate should \textsl{not} become completely
uniform, but should always vary in such a way that the global charge
vanishes.  For $N=2$ this requires a pair of domain walls, separating
the regions with positive and negative charge density.  Such domain
walls were observed and characterized in \cite{Gasenzer:2011by}.  For
$N>2$ there are several charges, and the charge density can revolve
smoothly through the $N(N-1)/2$ dimensional space of possibilities.

\begin{figure}[ht]
\putbox{0.32}{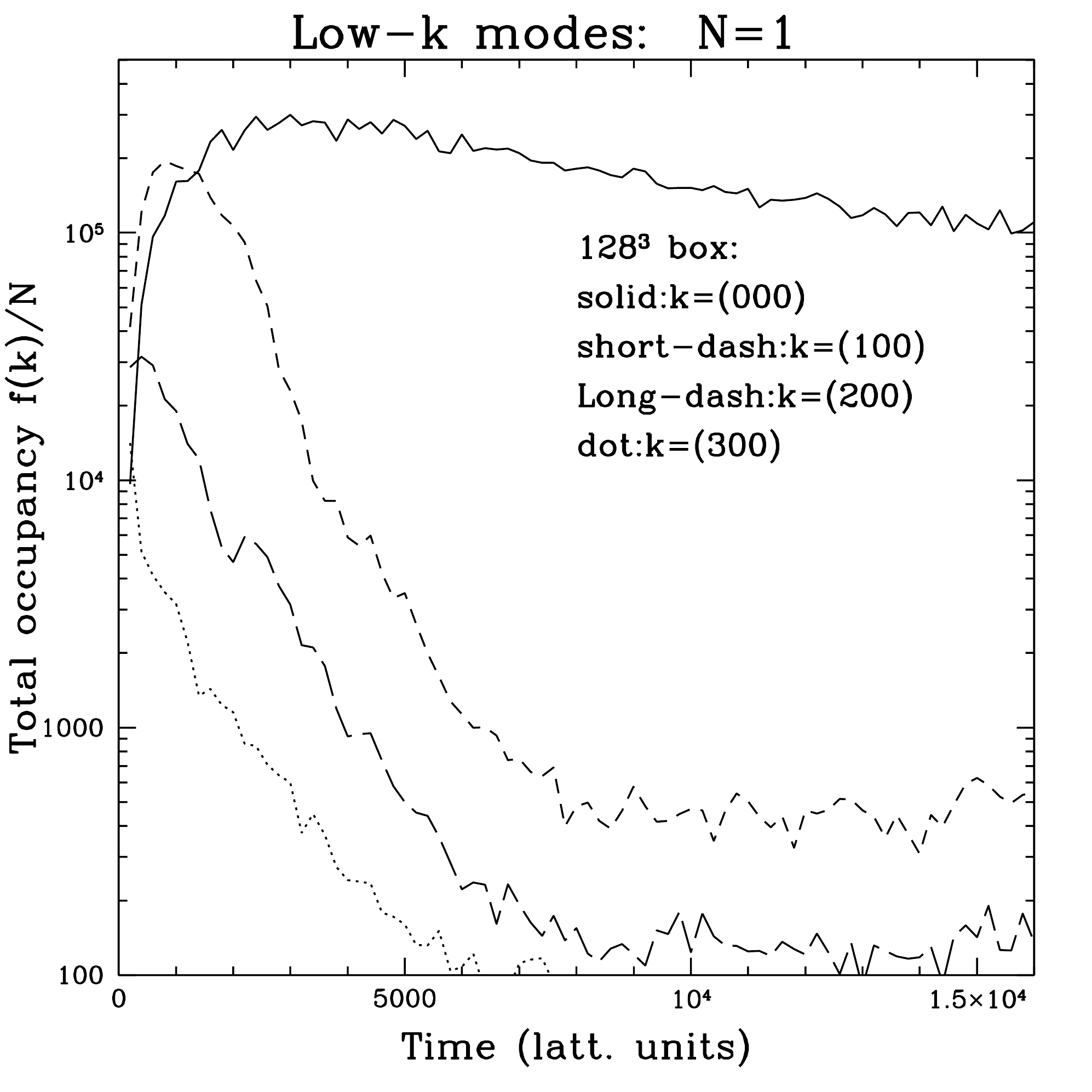}  \hfill
\putbox{0.32}{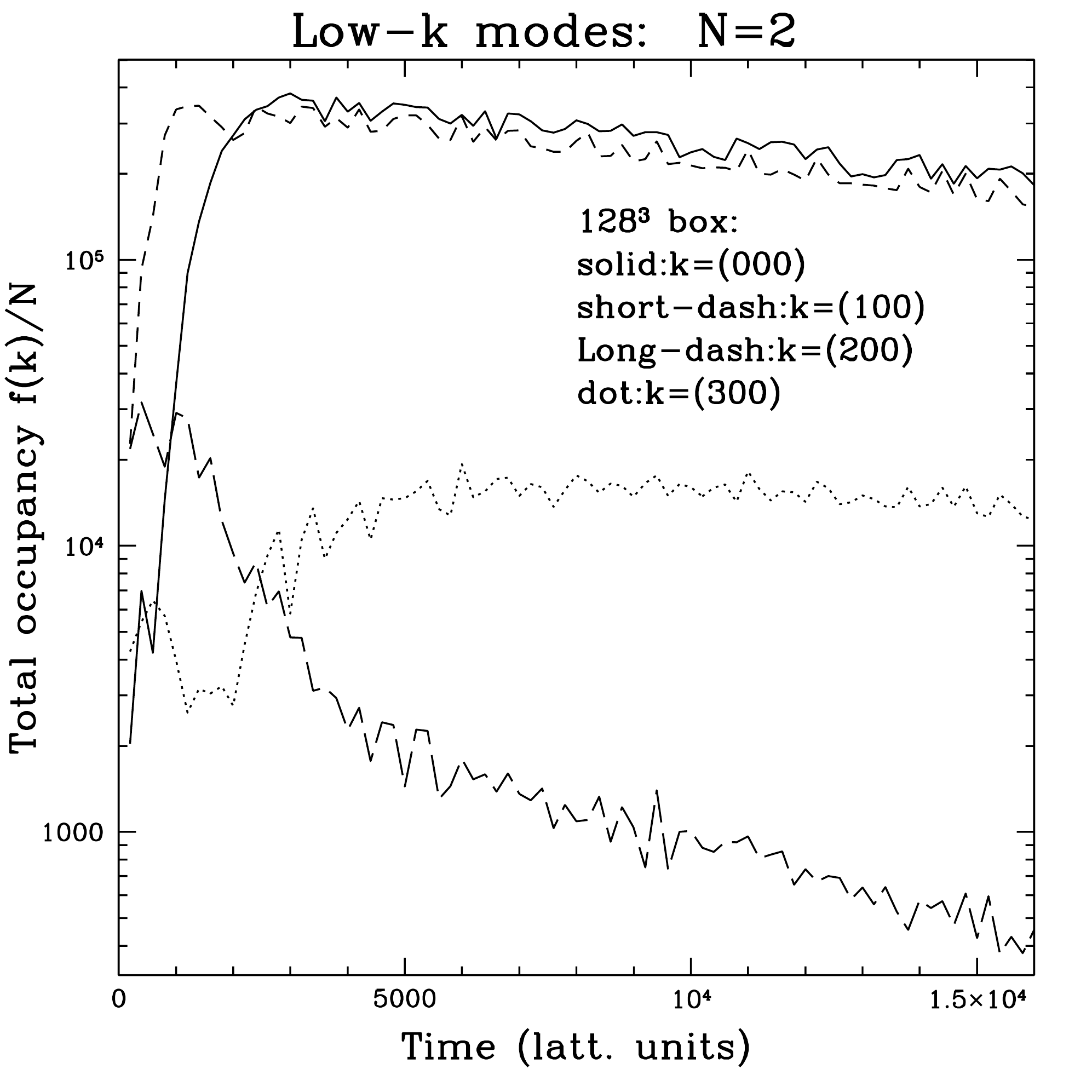}  \hfill
\putbox{0.32}{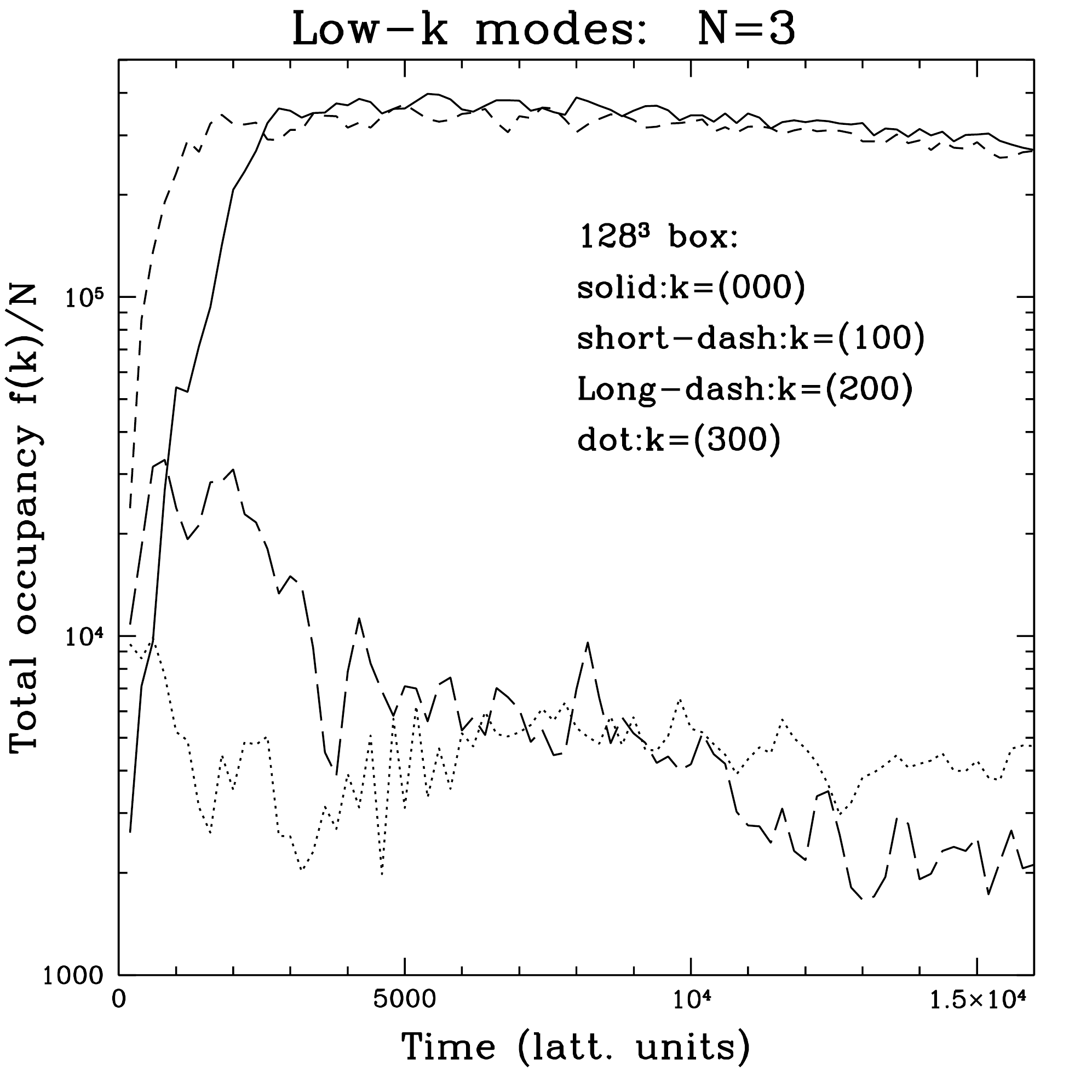}  \hfill
  \caption{\label{k0123}
    Power in the lowest $k$-modes which point along a lattice
    direction, for $N=1,2,3$ (left to right).  For $N=1$ the power
    all concentrates into the lowest mode.  For the other cases, the
    power is shared equally between the $k=0$ mode and the next-lowest
    mode; for $N=2$ there is also power in the $k=(3,0,0)$ mode.}
\end{figure}

First we look for this behavior in Fourier space, by evolving
$N=1,2,3$ systems in $128^3$ boxes, which allows us to achieve times
where the condensate completes its evolution towards the infrared.
\Fig{k0123} shows the total power in the $k=\frac{2\pi}{L}(0,0,0)$
mode, and the sum of power in
$k=\frac{2\pi}{L}(\pm l,0,0)$, $\frac{2\pi}{L}(0,\pm l,0)$
and $\frac{2\pi}{L}(0,0,\pm l)$ for 
$l=1,2,3$, for a typical simulation.  That is, it shows the total
power in each of the four
lowest Fourier modes for which $k$ lies purely along a lattice
direction.  For the case $N=1$ we see that the power all moves into
the $k=0$ mode, which then decays with time.  For $N=3$ the power is
shared equally between the $k=0$ mode and the modes of form
$k=(\pm 1,0,0)\frac{2\pi}{L}$.  It is also shared equally between
$\phi^2$ an $\dot{\phi}^2$, in that the ratio
$\langle \dot\phi^2\rangle / \langle \phi^2 \rangle$ summed over the
lowest modes remains fixed, rather than oscillating as it does for
$N=1$.  The case $N=2$ is superficially similar to the $N=3$ case,
except that the power in $l=3$ does not decay but remains much larger
than the power in $l=2$ (or any of the other Fourier modes, such as
$k=\frac{2\pi}{L}(1,1,0)$, which are not along lattice directions).
This is because, for $N=2$, the condensate is discontinuous across a
domain wall, which leaves power in all odd harmonics, while for
$N=3$ the charge density is smooth.  We also studied $N=4$, which
shows the same behavior as $N=3$.

\begin{figure}[htb]
  \hfill \putbox{0.4}{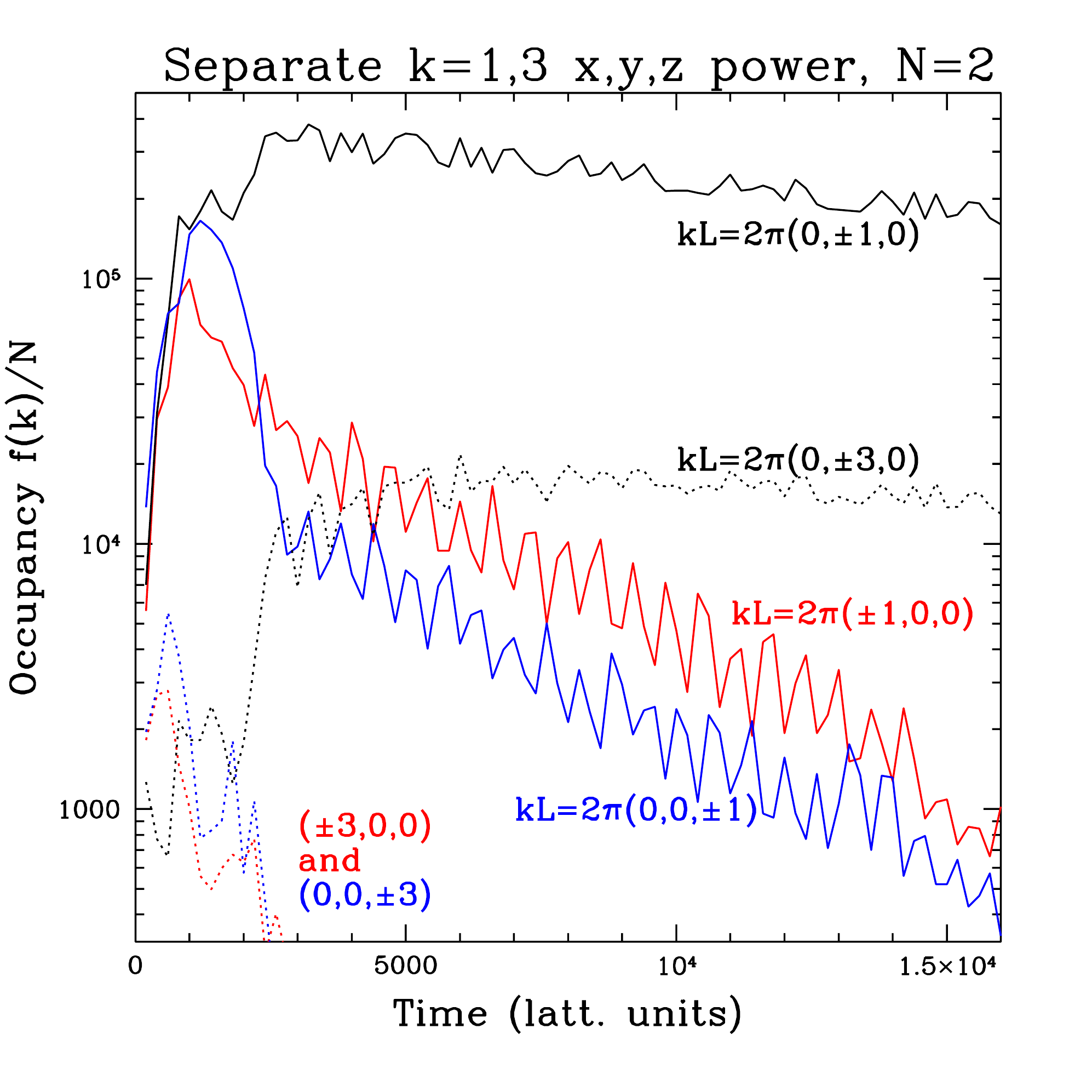}
  \hfill \putbox{0.4}{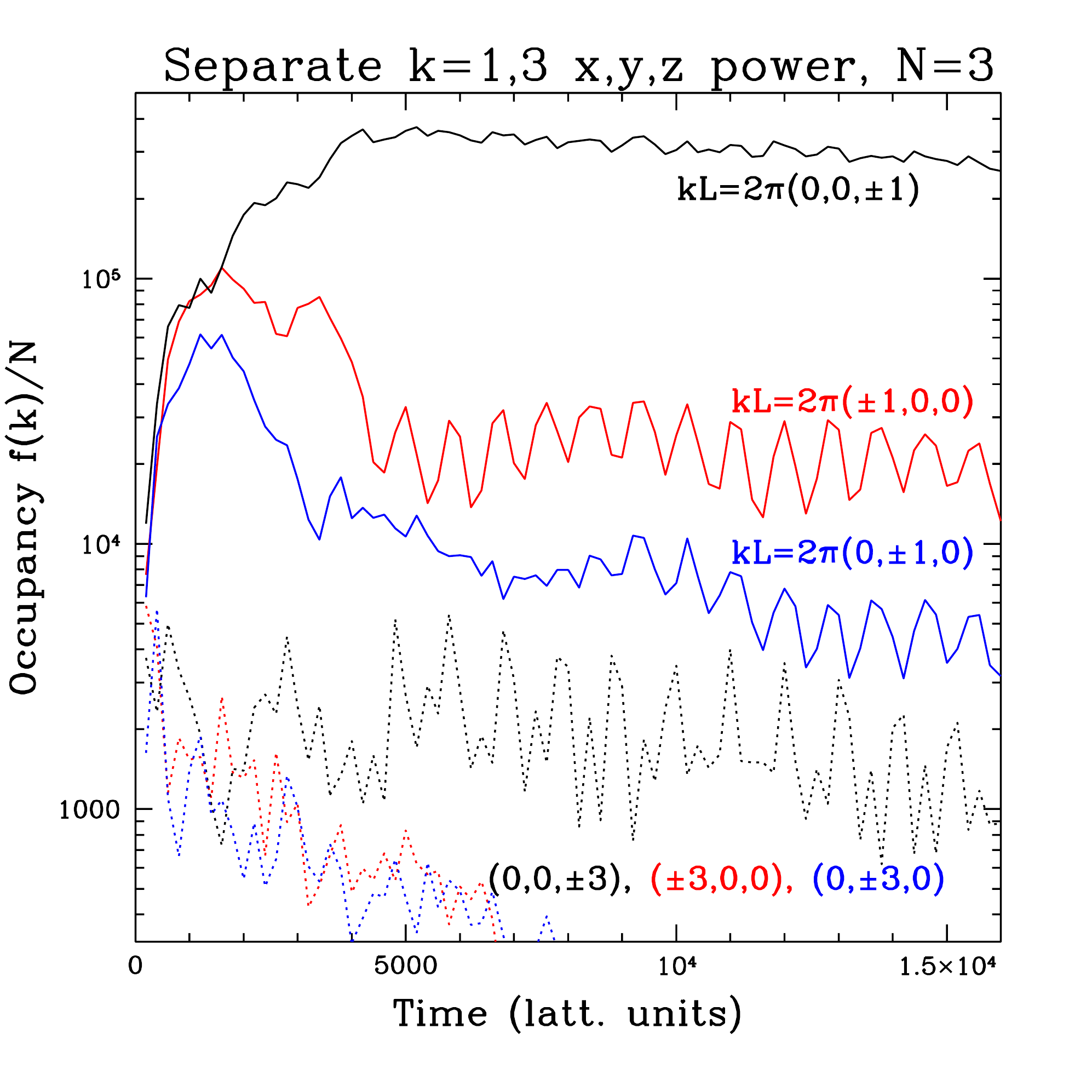}
  \hfill $\vphantom{.}$
  \caption{\label{kxyz}
    Power in the lowest nonzero Fourier mode along the $x$, the $y$,
    and the $z$ axis, as well as the $(3,0,0)$-type modes.  Left:
    $N=2$.  Right:  $N=3$.  In each case, the power along two axes
    dies away, while the third axis persists.  For $N=2$ it is the
    same axis where there is persistent power in the $(0,3,0)$ modes.}
\end{figure}

What \Fig{k0123} does not show is that the power in $(1,0,0)$ type
modes is in fact all concentrated in the mode along one axis.  To see
this, in \Fig{kxyz} we plot separately the power in
$k=\frac{2\pi}{L}(\pm 1,0,0)$ and that in
$k=\frac{2\pi}{L}(0,\pm 1,0)$ and $(0,0,\pm 1)$, for $N=2$ and $N=3$.
We also plot the $(3,0,0)$ power, to see that it is supported along
the same lattice direction as the $(1,0,0)$ power for the $N=2$ case
where it does not die off.  The fact that the power
ended up along one axis indicates that the condensate remains
spatially asymmetric in this, but not the other two, lattice
directions.  We performed each simulation several times with different
random number seeds, and found that the lattice axis where the
breaking occurs is randomly different, but the pattern shown above is
the same for each simulation; for $N>1$ the power equipartitions
between the $(0,0,0)$ mode and one set of $(\pm 1,0,0)$ modes and is
small in the others.

\begin{figure}[htb]
  \hfill
  \putbox{0.4}{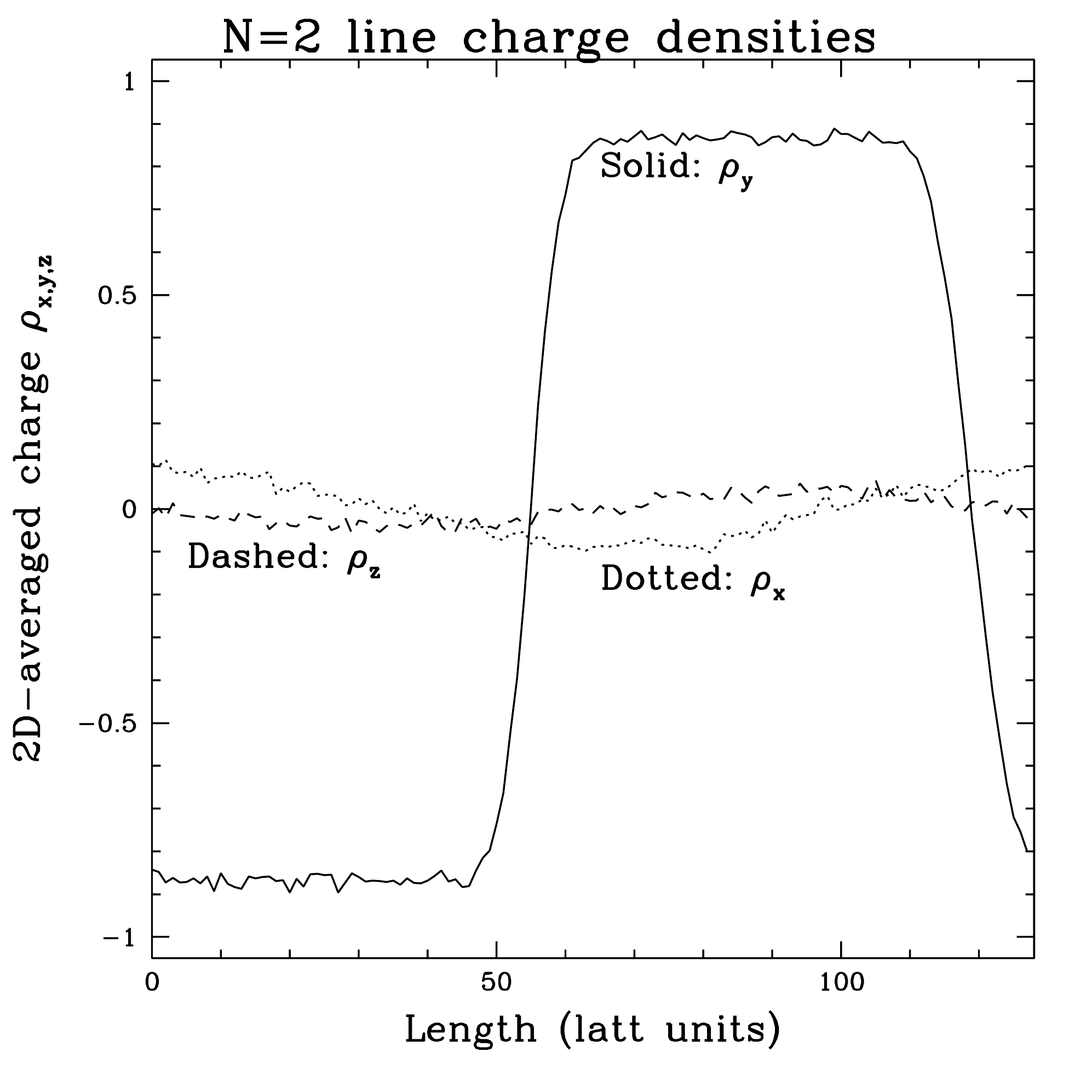} \hfill
  \putbox{0.4}{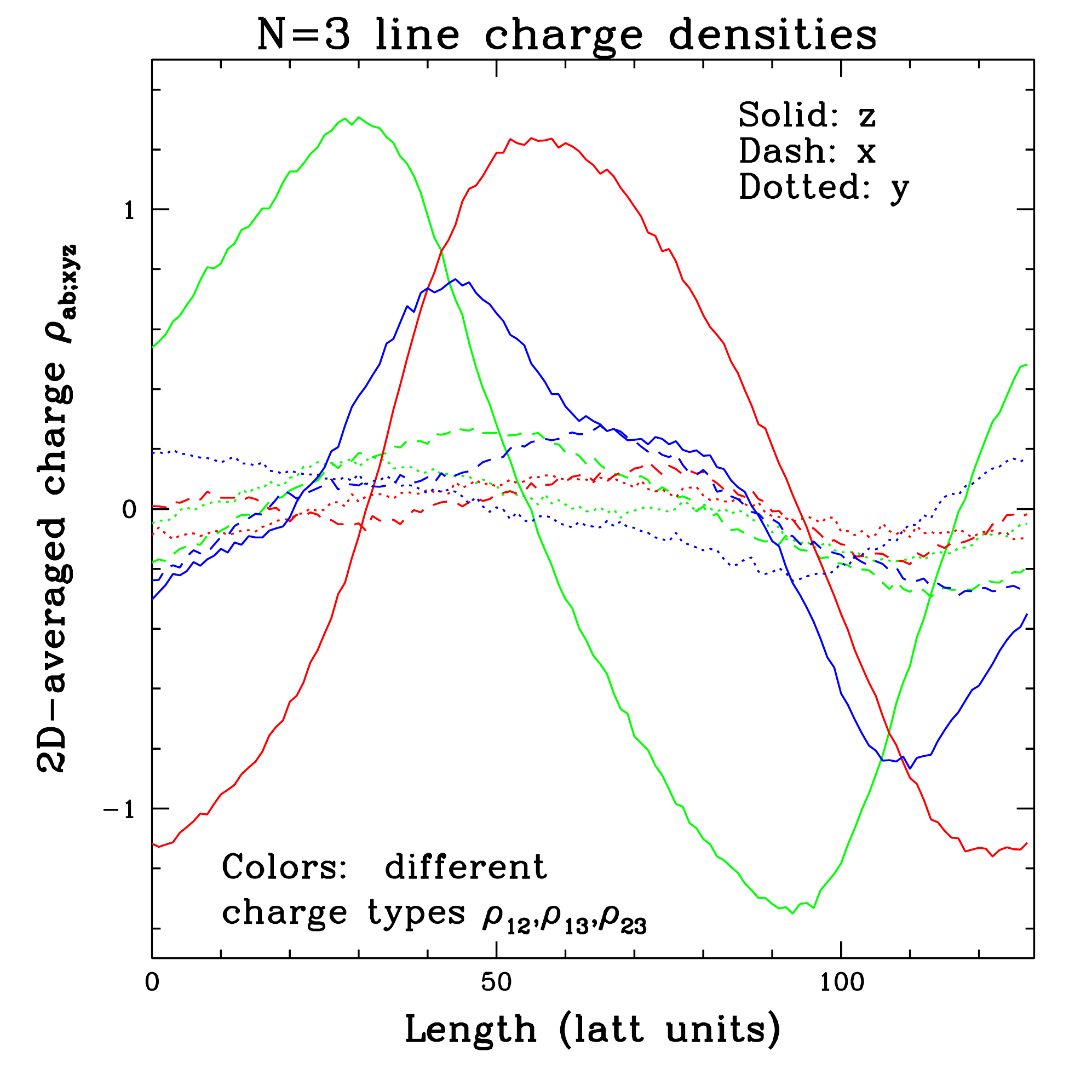} \hfill $\vphantom{.}$
  \caption{\label{rhoxyz}
    plane-averaged charge density plotted along the remaining axis,
    in $128^3$ box at $t=10,000$ when the condensate has settled into
    its final form.  Left: $N=2$ case, where the charge density varies
    along the $y$ axis. Right: $N=3$ case, where the three charge
    densities vary primarily in the $z$ direction.}
\end{figure}

To see it another way, let us define the charge density averaged
over two lattice directions as a function of the third:
$\rho_{ab;x} \equiv L^{-2} \int dy\,dz\: \rho_{ab}(x,y,z)$ and similarly for
$\rho_{ab;y}$ and $\rho_{ab;z}$.  We plot this, for each axis and at a
fixed late time $t=10,000$ lattice units, in \Fig{rhoxyz}.  On the
left, we see that the $N=2$ theory has a pair of sharp discontinuities
in $\rho_{y}$, with plateaux between.  For $N=3$ we see that each
charge component varies smoothly such that
$\sum_{ab} \rho^2_{ab;z}$ is nearly constant and much larger than in
the other two directions.  Note that these measurements were made
directly on the fields without any smearing and no averaging except
for the averaging over lattice planes.  Averaging over 2D planes is
already enough to suppress the contribution from UV modes, without any
field smearing.

\begin{figure}[ht]
  \centerline{\putbox{0.45}{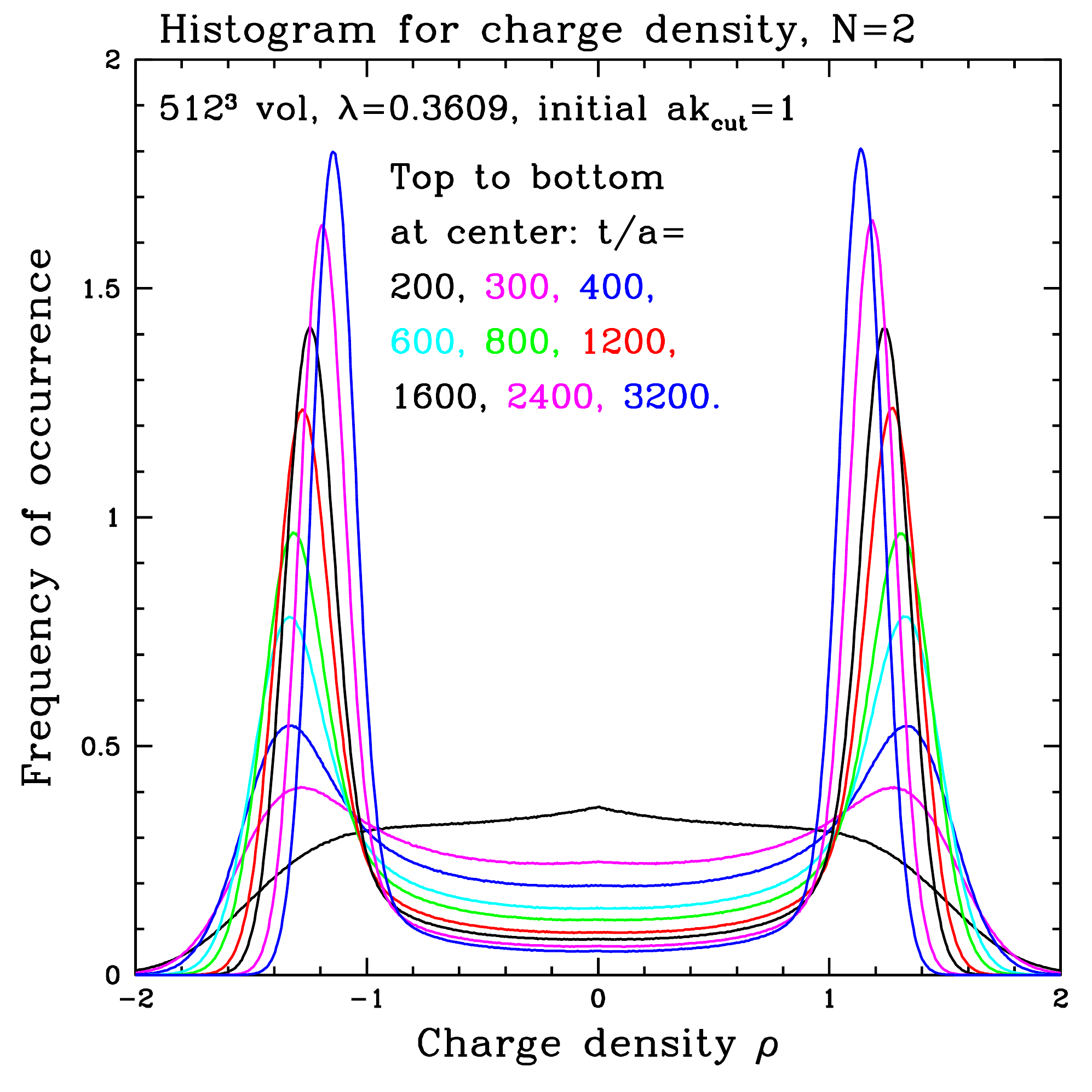}}
  \caption{\label{rhohist}
    Histogram of the condensate's local charge density, for the $N=2$
    theory in a $512^3$ box after smearing $\phi$ and $\dot\phi$ to
    eliminate the contribution of UV fluctuations.}
\end{figure}

We can also form a histogram of the value of $\rho$ at each point on
the lattice.  The result looks fairly Gaussian, because of the large
site-by-site contribution from UV fluctuations.  But if we first
smooth the fields, the result is different.  Sticking with the $N=2$
theory, we Fourier transformed $\phi$ and $\dot\phi$, multiplied by
$\phi(k) \to \phi(k) \exp(-k^2/\omega^2)$,
$\dot\phi(k) \to \dot\phi(k) \exp(-k^2/\omega^2)$ (with $\omega$
determined from the data as described in Appendix \ref{lattdetail}),
and Fourier transformed back, in order to eliminate the contribution
of UV fluctuations.  Then we determined the charge density due to IR
fields at each lattice point by finding
$\rho(x)=i(\phi^* \dot\phi(x) - \phi\dot\phi^*(x))$ as usual.
\Fig{rhohist} shows a histogram of how common
each possible value of the charge density is, for several times and in
a $512^3$ box.  The curves are normalized so the area under each curve
is 1.  At late times there are two sharp peaks with a plateau
in between; we interpret the peaks as arising from points within
regions of one or the other charge phase, and the plateau as the
contribution from the domain walls separating these phases.

\subsection{Ordering dynamics}
\label{secorder}

\begin{figure}[ht]
  \putbox{0.32}{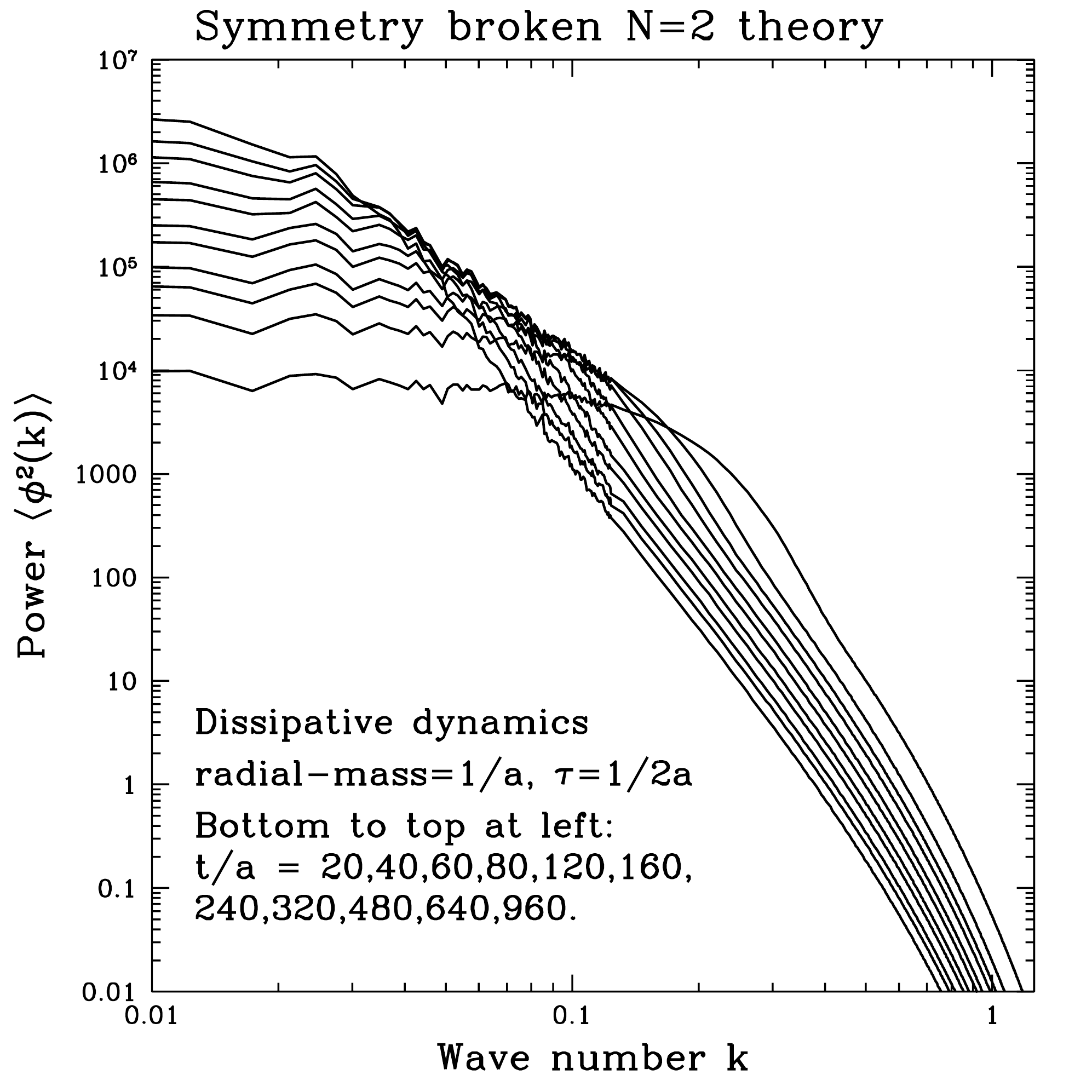} \hfill
  \putbox{0.32}{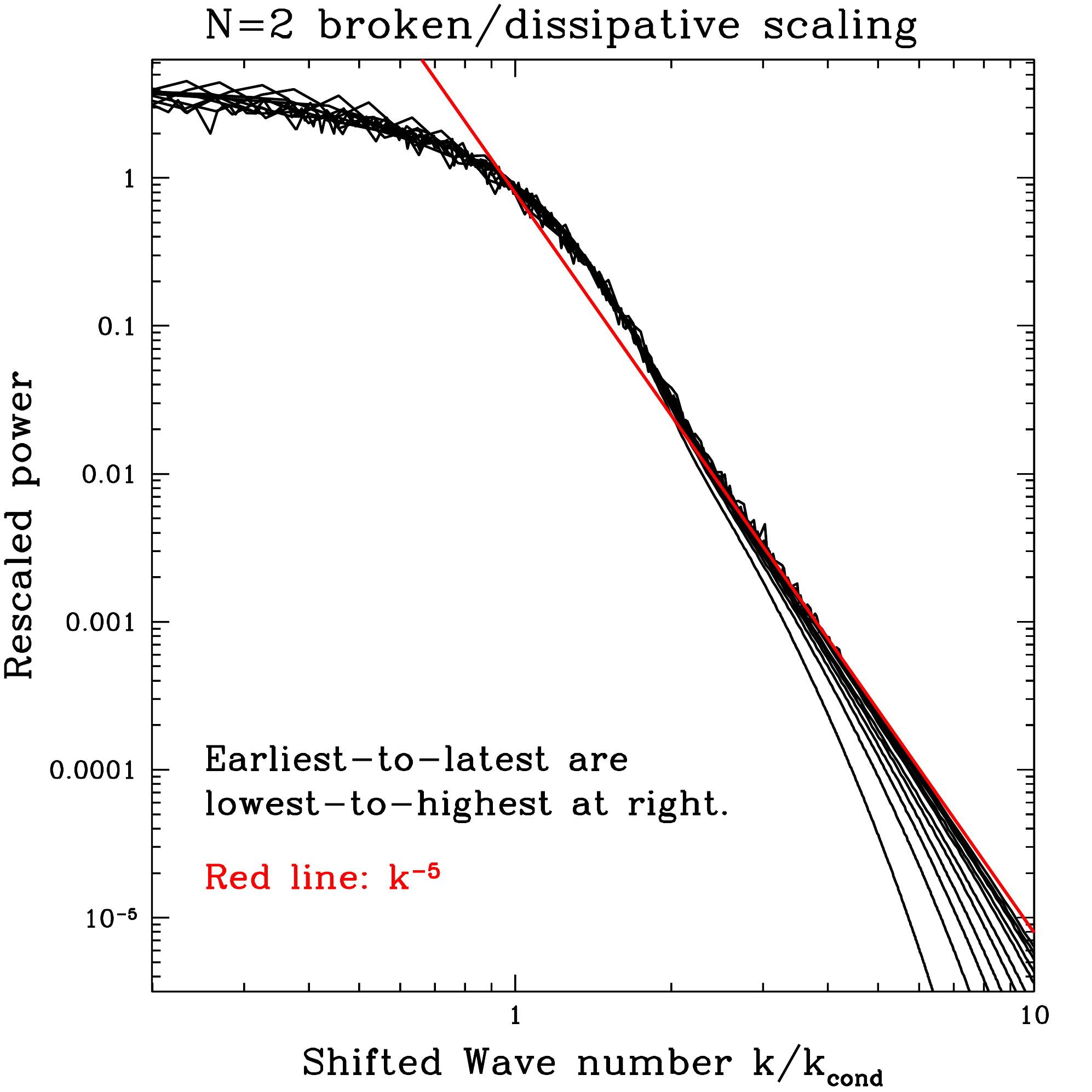} \hfill
  \putbox{0.32}{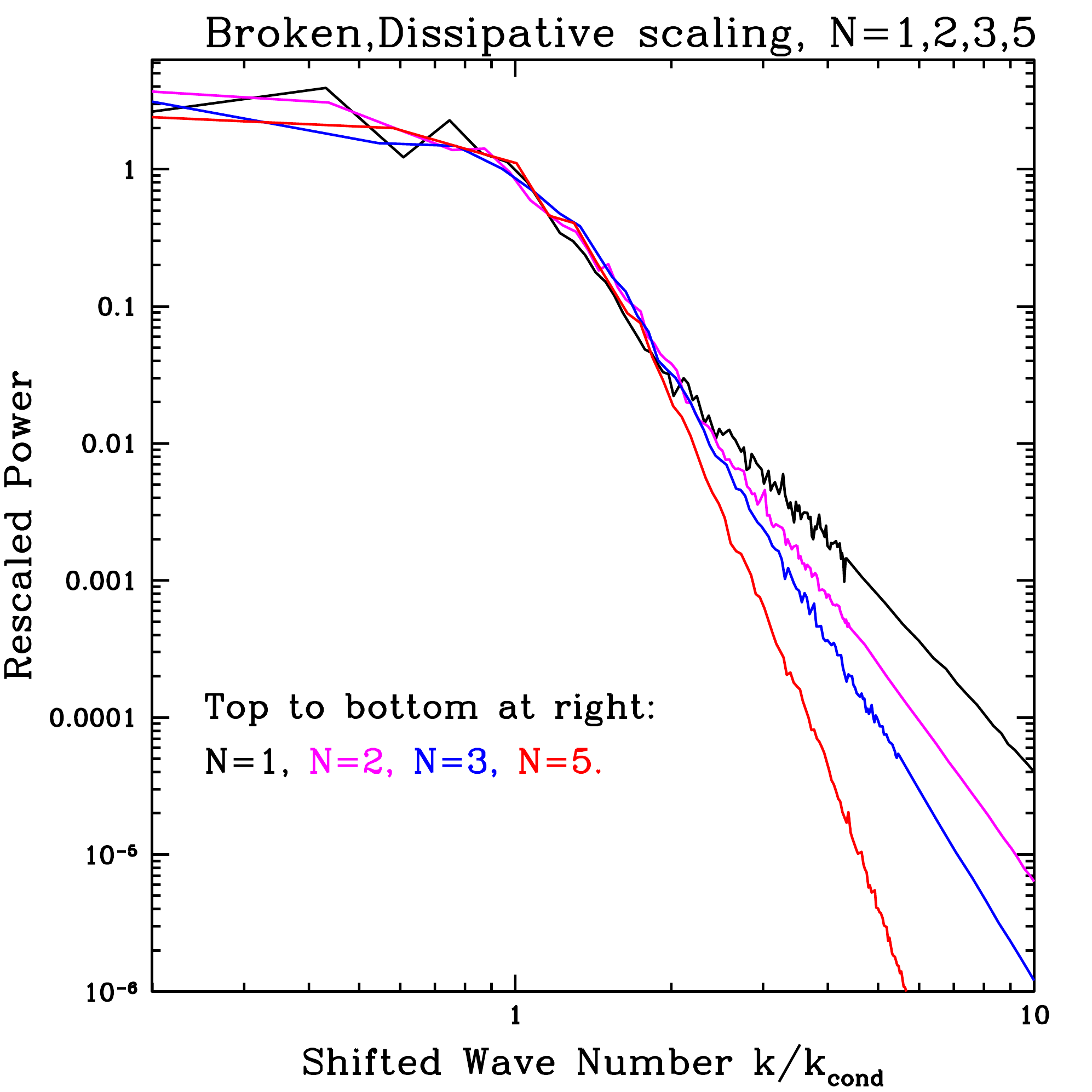}
  \caption{\label{decayfig}
    Power spectrum for broken-symmetry theory under dissipative
    dynamics, to illustrate the effect of defects on the spectrum.
    Left: $N=2$ theory at several times.  Middle:
    rescaling $k\to k/\kcond$ and power by $2\pi^2\kcond^{-3}$ so the
    curves collapse onto a scaling form.  Right:  scaling form for
    $N=1,2,3,5$ superposed, showing different power-law tails.}
\end{figure}

Next consider a large lattice volume $512^3$ and somewhat shorter
times, which allows us to see the field organization dynamics of the
local condensate.  As we have emphasized, for $N\leq 4$ the condensate
can carry topological defects.  A
topological defect is a location where the condensate field varies
abruptly.  Therefore it is guaranteed to create power-law tails in the
high-$k$ part of the power spectrum for the condensate.  Specifically,
a network of domain walls should create $k^{-4}$ tails in the power
spectrum; strings should create $k^{-5}$ tails, and monopoles should
create $k^{-6}$ tails.%
\footnote{In general the power is $k^{-d-c}$ with $d$ the space
  dimension and $c$ the codimension of the defect.  Consider a defect
  with codimension $c$ and dimension $d-c$.  A function in $c$
  dimensions which is discontinuous at a point has a Fourier spectrum
  with $k^{-2c}$ tails; but each of the remaining
  $d-c$ dimensions ``along'' the defect dilute this by an additional
  power.}
To illustrate this behavior, consider \Eq{Lagrangian} with $m^2<0$ and
adding a damping term to the dynamics such that
\be
\mbox{Dissipative dynamics:} \quad
\partial_t^2 \phi_a = -\tau \partial_t \phi_a
+ \nabla^2 \phi_a + V_{,\phi_a} \,.
\ee
In our numerical implementation we chose $\tau = a/2$ so the damping
dynamics are strong, and we chose $a^2m^2=-0.5$ and $\lambda=1$ so the VEV
is $|\phi|=1$ in lattice units and the radial fluctuations have
oscillation frequency
$1/a$.  In this case the field forms a true condensate, but we choose
random initial conditions, so the field starts out disordered and must
follow ordering dynamics.  The ordering dynamics are impeded by
topological defects with codimension $N$: walls for $N=1$, strings for
$N=2$, monopoles for $N=3$, textures for $N=4$, but no topological
structures for $N=5$.  By the arguments above, we expect the power
spectrum for $N=1,2,3$ to display $k^{-(3+N)}$ tails, and indeed it does.
\Fig{decayfig} shows the power spectrum at several times for the case
$N=2$, first with fixed axes and then with axes rescaled such that the
curves collapse onto a single scaling behavior.  Finally, the scaling
behavior is compared for $N=1,2,3,5$ (at time $t=960a$).  The IR
behavior after scaling
to $\kcond$ looks similar for all cases, but the larger-$k$ behavior
is different.  For $N=1,2,3$ we see power laws with $k^{-3-N}$ power,
and for $N=5$ the tail does not asymptote to a power, all as
expected.  Note that for $N=1,2$ the spectrum is actually steeper than
$k^{-3-N}$ around $k=1.8\kcond$, before taking the power-law form
at $k>2\kcond$.

We also see in the two left plots of \Fig{decayfig} that the power-law
tails break off at a larger-$k$ scale which stays fixed in physical
units.  This scale is set by the thickness of the defects; at this
scale the spatial features become smooth so the power spectrum changes
from power to exponential behavior.  Again we emphasize that this
discussion is only to show the effect of defects on the power-law
tails.

Returning to the theory with a particle-number condensate, we expect
the topological structures to be \textsl{one} contribution to
the IR tail of the condensate's part of the power spectrum.
There can also be ordinary nontopological fluctuations in the
condensate, which will obscure the topological contribution if they
have a softer power law or substantially larger amplitude.  We also
expect the defect core size to depend on $N$, now differing between
the $N=2$ case and other cases.  For string or monopole defects we
expect the core size to be $\sim \omega^{-1}$.  But for domain walls,
we argue that it should be at least 3 times larger.  Generally the
thickness of a domain wall is set by the potential energy cost of the
state at the center of the wall, with inverse width
$k_{\mathrm{wall}} \propto \sqrt{V}$.  The core of the domain wall
contains condensate which is not locally charged but oscillates
straight back-and-forth.  The excess energy cost of this state is only
about $13\%$ of the condensate's energy density, so we expect the
wall's inverse thickness to have an extra power of $\sqrt{0.13}$.

\begin{figure}
  \hfill \putbox{0.45}{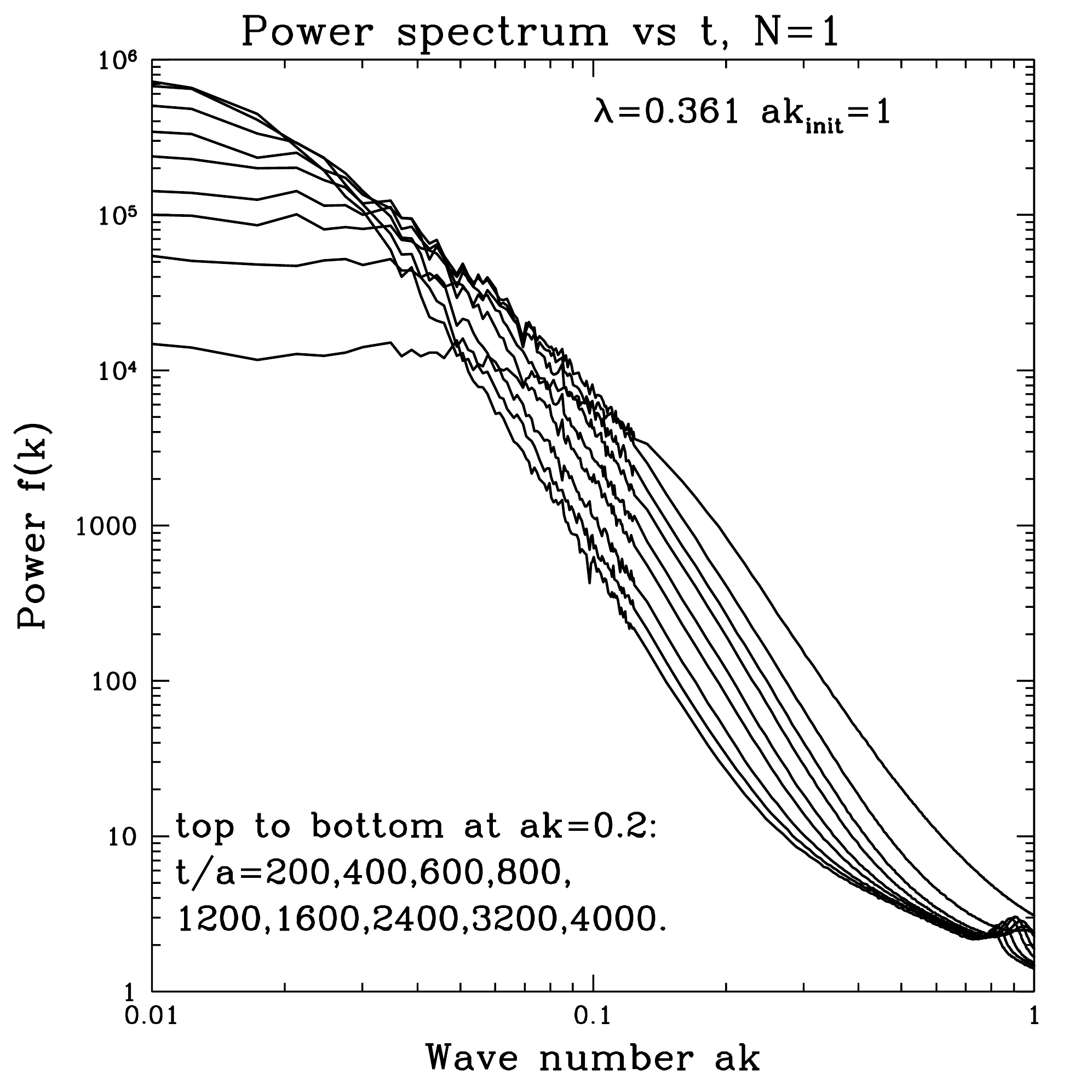} \hfill
  \putbox{0.45}{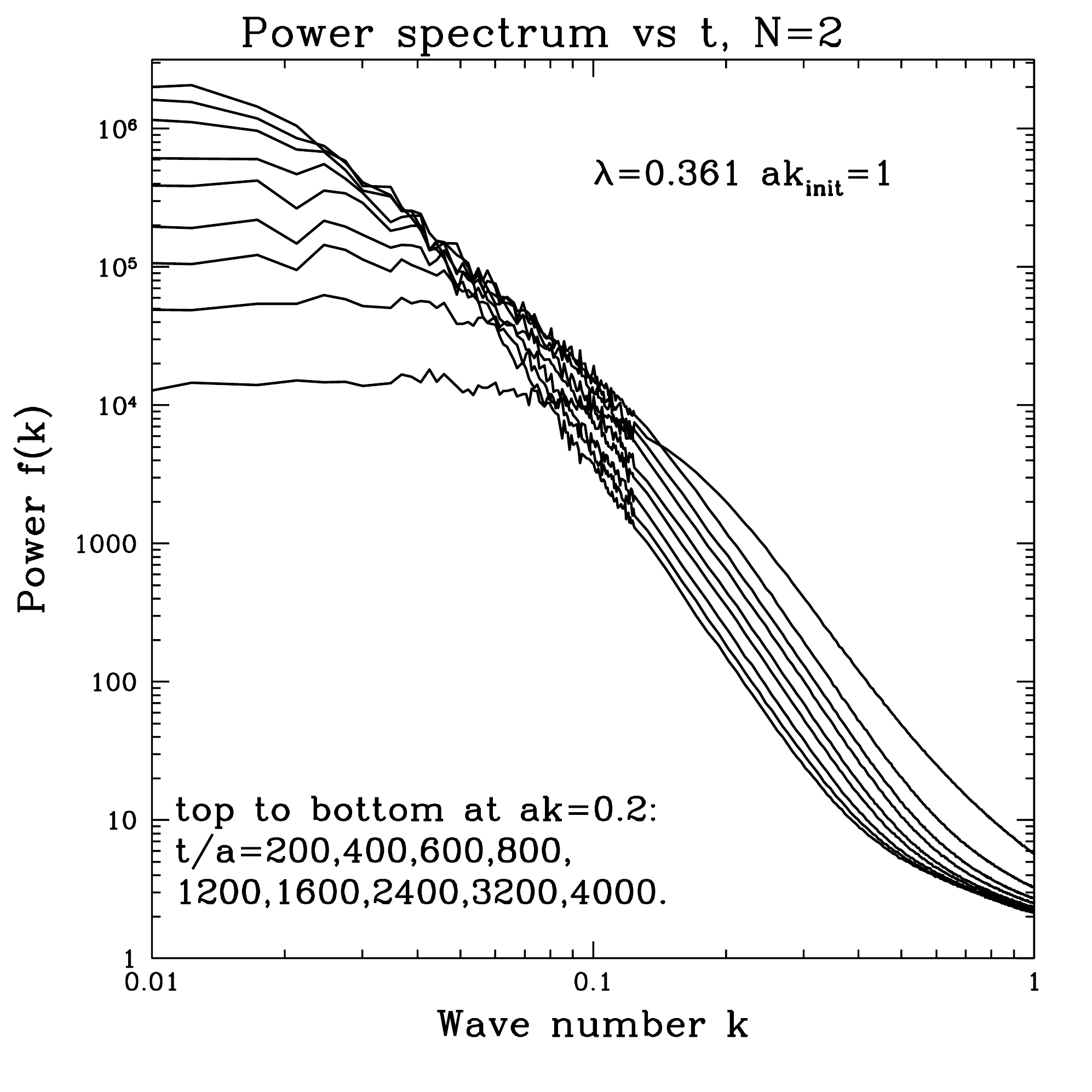} \hfill $\phantom{.}$ \\
  \hfill \putbox{0.45}{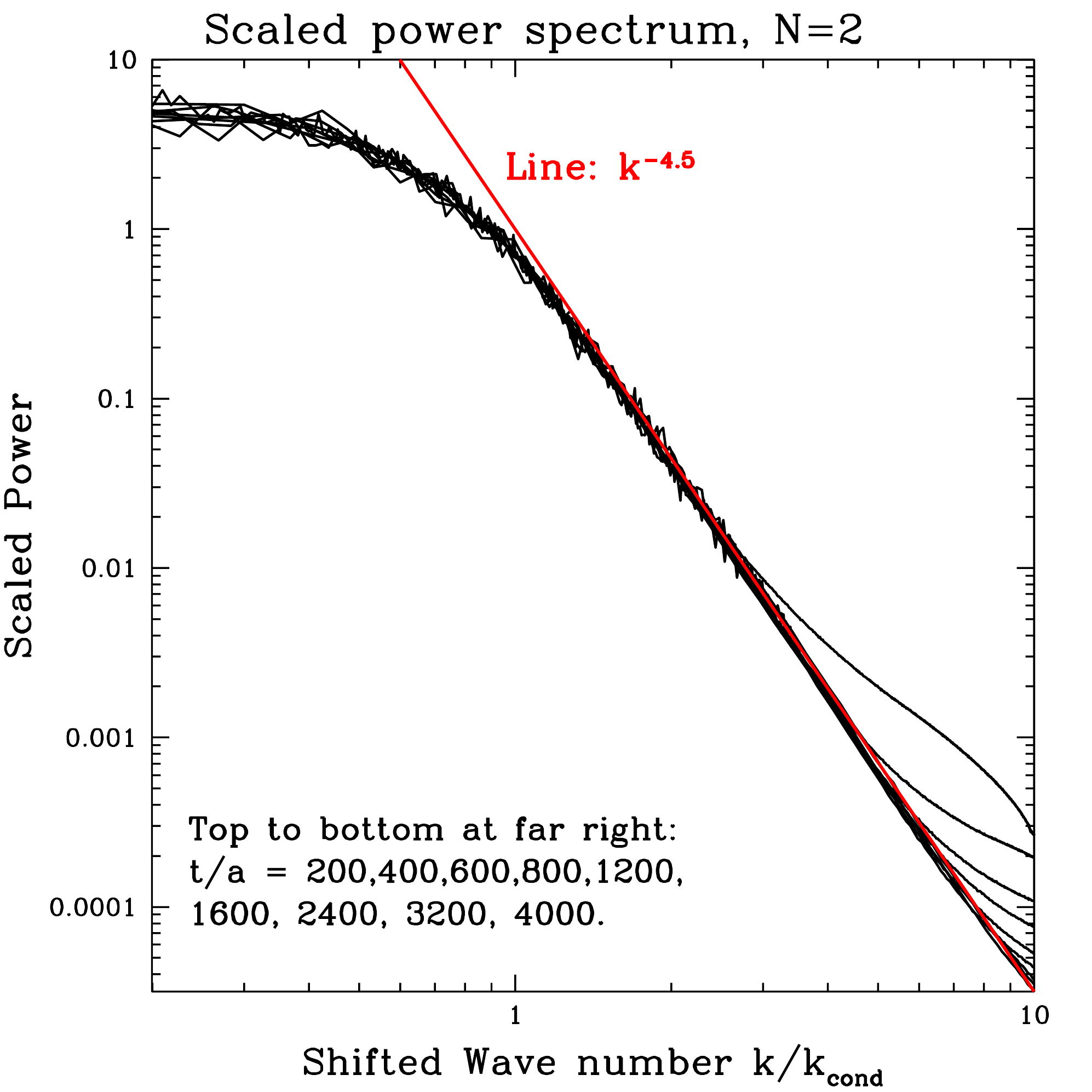} \hfill
  \putbox{0.45}{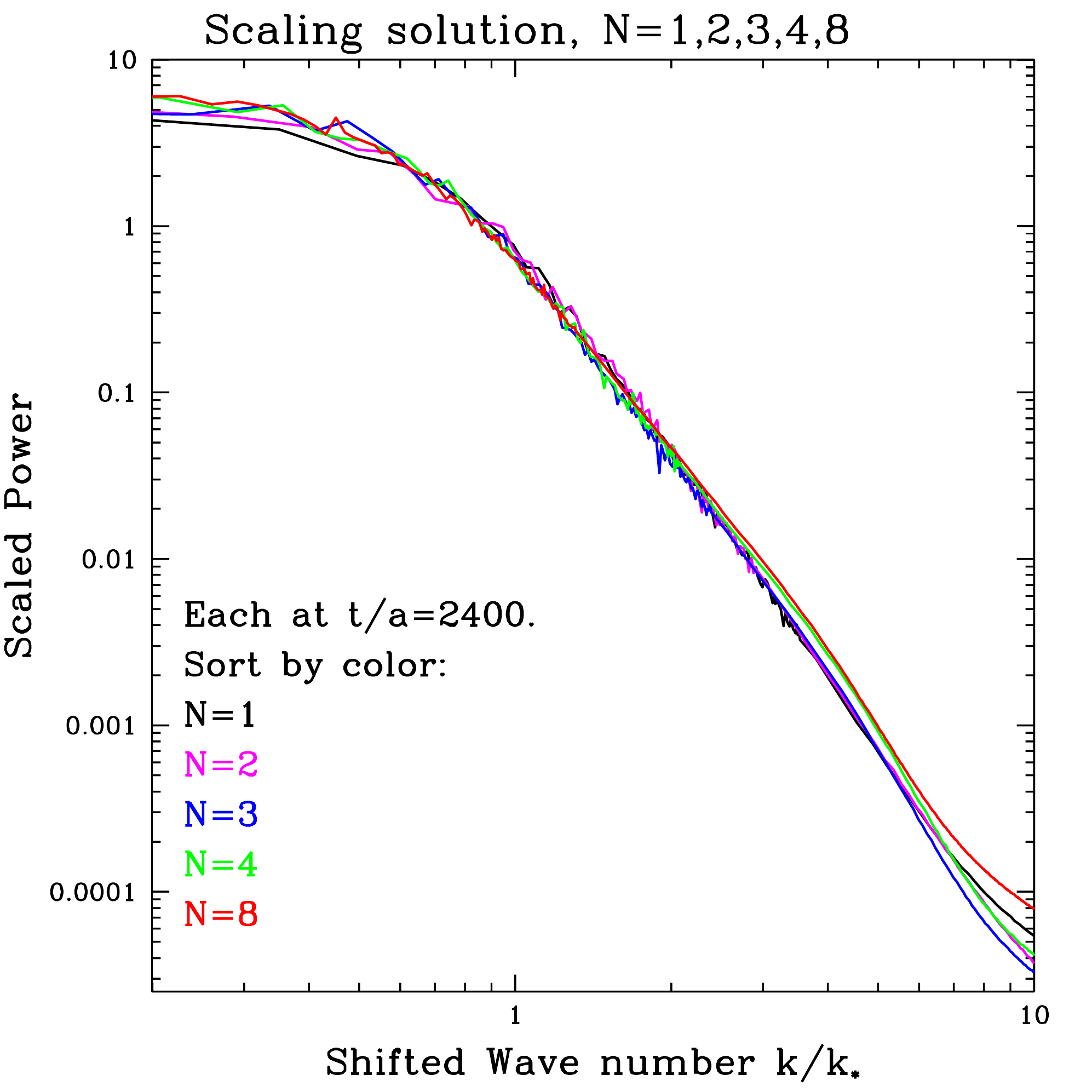} \hfill $\phantom{.}$
  \caption{\label{power}
    Power spectrum of the IR condensate.  Top:  power spectrum at
    several times for $N=1$ theory (left) and $N=2$ theory (right).
    Bottom left:  rescaled version for $N=2$ showing scaling
    behavior.  Bottom right:  scaled spectrum at $t=2400$ for
    $N=1,2,3,4,8$.}
\end{figure}

\Fig{power} shows the analogous figures to \Fig{decayfig}, but for the
theory with a condensate.  The time-dependence of the spectra in the
infrared is superficially similar to that in \Fig{decayfig}.
As expected, this behavior flattens out at larger $k$ rather than
getting steeper.  In the $N=1$ case, but not for any other $N$, there
is also a feature around $ak=0.9$.  This peak is the freshly-produced
excitations of frequency $2\omega_0$, generated from the decay of the
condensate, see \cite{berg2} and the next subsection.

\Fig{power} also shows the spectra rescaled to have the same total
power and characteristic wave-number $\kcond$.  It shows excellent
collapse onto a scaling solution, broken at a larger-$k$ scale which
grows, in units of $\kcond$, as time progresses.%
\footnote{The physical scale decreases, as we see in the upper plots;
  but $\kcond$ decreases faster.}
This collapse onto a scaling function has already been
observed by previous authors \cite{Berges:2010ez,Orioli:2015dxa}.
In \cite{Berges:2010ez} it is argued that the steep part
of the spectrum below $\kcond$ should behave as $f\propto k^{-5}$.  We
find $k^{-4.5}$, which is the slope of the straight line in the
lower-left frame of the figure, which agrees with the numerical
results in \cite{Orioli:2015dxa}.  A $k^{-5}$ line visibly fails to
fit (not shown).

Finally, in the lower right we show the scaling solutions for
$N=1,2,3,4,8$ all superposed.  Beyond about $k=7\kcond$ the curves
deviate from scaling, so this behavior should be ignored.  The curves
agree strikingly well, in strong contrast to the symmetry-broken case
of \Fig{decayfig}.  This collapse of different times and $N$ values
onto a single scaling solution has previously been pointed out by
\cite{Berges:2010ez,Orioli:2015dxa}.  These authors argued that the
concordance indicates a universality in the physical origin of the
spectrum.   This is rather surprising, given the differences in
condensate structure and topological obstructions which we have
previously discussed.  Therefore it does
\textsl{not} appear that topological structures control the tail's
behavior, at least for the cases other than $N=2$.

It is a little puzzling that a $k^{-4}$ tail does not emerge for the
$N=2$ case, since we have already seen that there are domain walls in
the charge-sign of the condensate.  One possibility is that a $k^{-4}$
region would emerge if we could achieve a much wider scaling window.
After all, for the symmetry-breaking theory in \Fig{decayfig}, the
scaling emerges relatively late, and as discussed above we expect it
to break down by $k\sim \omega/3$ because the walls are quite thick.
Unfortunately, achieving a significantly wider scaling window is
prohibitively numerically costly.

\subsection{Condensate decay}
\label{secdecay}

\begin{figure}[tb]
\centerline{\putbox{0.6}{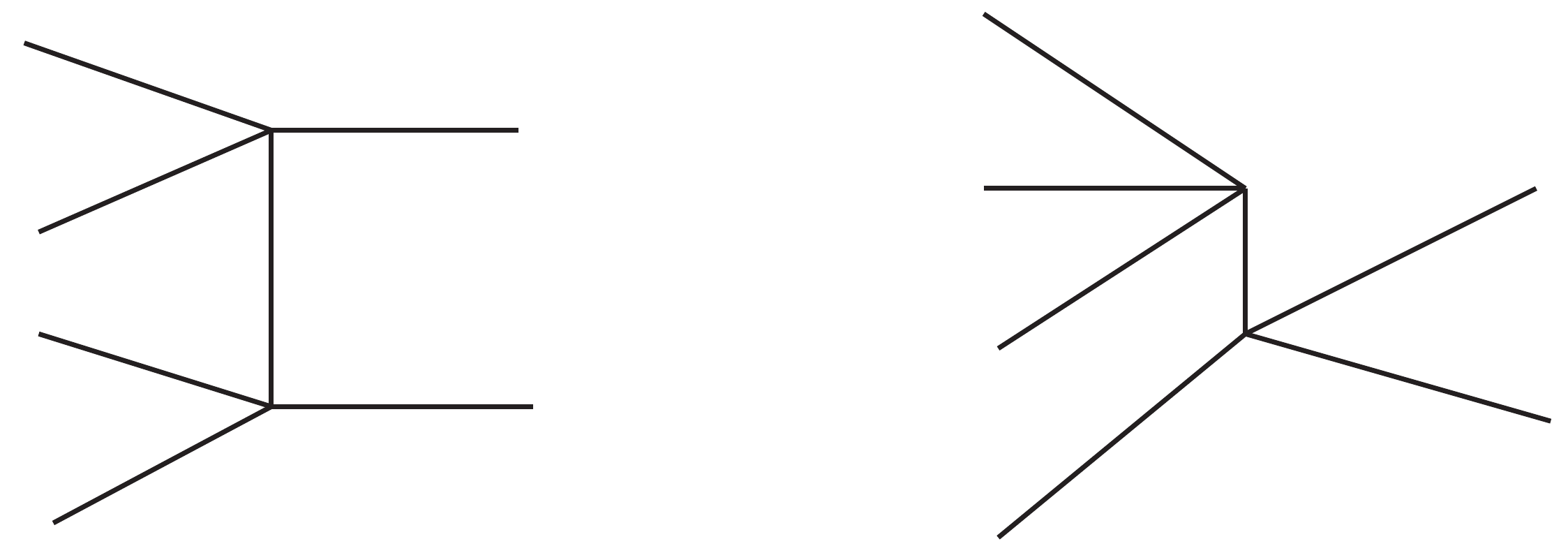}}
\caption{\label{42diagrams}
  Feynman diagrams for $4\to 2$ number-changing scattering.}
\end{figure}

The presence of a large local charge density has consequences for how
the condensate decays with time.  To see this, consider the Feynman
diagrams responsible for condensate decay; $4\to 2$ scattering as
shown in \Fig{42diagrams}.  For the $N=1$ theory, the leading cause of
condensate decay is the case where the 4 incoming particles are drawn
from the condensate and the two outgoing particles are finite-$k$
excitations, each carrying $2\omega$ energy.  The two diagrams
partially cancel, since one has a spacelike propagator and one has a
timelike propagator above the mass shell.

It is actually not
straightforward to compute the diagram, since the dispersion relation
for the outgoing particles and virtual propagators are not simply
those of a massive particle -- the coherence of the condensate is not
the same as $\langle \phi^2 \rangle$ arising from many independent
fluctuations.  But we can parametrically estimate the rate fairly
easily.  Though the calculation can be done within the classical
theory, we will use the notation of the quantum theory because it is
probably more familiar to our readers.  The rate of particle number
change from the condensate is determined by the Boltzmann equation for
the condensate particles:
\bea
\label{decayest1}
\frac{d\ncond}{dt} & \sim &
\int \frac{d^3p_{1,2,3,4} d^3 k_{1,2}}{p_1 p_2 p_3 p_4 k_1 k_2}
|\mathcal{M}|^2 \;\delta^4\left( \sum p_i - \sum k_i \right)
\nonumber \\ && \qquad \times
\left( f_{p_1}f_{p_2}f_{p_3}f_{p_4} [1{+}f_{k_1}][1{+}f_{k_2}]
 - [1{+}f_{p_1}][1{+}f_{p_2}][1{+}f_{p_3}][1{+}f_{p_4}] f_{k_1}
 f_{k_2} \right)
\nonumber \\ & \sim &
\int \frac{d^3p_{1,2,3,4}}{p_1 p_2 p_3 p_4} f_{p_1}f_{p_2}f_{p_3}f_{p_4}
   \frac{d^3 k_{1,2}}{k_1 k_2} (f_{k_1}+f_{k_2})
|\mathcal{M}|^2 \;\delta^4\left( \sum p_i - \sum k_i \right)
\nonumber \\ & \sim &
\lambda^4 \frac{\ncond^4}{\omega^8} f(2\omega) \,.
\eea
Here $\ncond = \int d^3 p f(p)$ is the particle number
density in the condensate, $\lambda^4$ arises from the squared matrix
element, and the powers of $\omega$ enter on dimensional grounds from
the remaining phase space integrals and the matrix element.

Naively this result seems to state that the condensate's decay rate is
a very steep power of the condensate's size $\ncond$.  But
the presence of $\omega^{-8}$ in the final expression moderates this,
because the dominant contribution to the oscillation frequency is
actually the condensate itself.  Indeed, if we ignore the contribution
from all other fluctuations, we have, from \Eq{N1omega2} and
\Eq{N1epsn},
\be
\label{omega3}
\omega \sim \lambda^{1/4} \varepsilon^{1/4} \quad \mbox{and} \quad
\varepsilon \sim \lambda^{1/3} n^{4/3} \quad \Rightarrow \quad
\omega^3 \sim \lambda \ncond \,.
\ee
Therefore we expect weaker $\ncond$ dependence:
$d\ncond/dt \propto \ncond^{4/3} f(2\omega)$.  If, for instance,
$f(2\omega) \propto \omega^{-3/2}$, this would give
$d\ncond/dt \propto \ncond^{5/6}$.  In general $f(2\omega)$ also
varies (decreases) with time. And
any other contribution to the oscillation frequency, such as
an explicit $m^2$ term or a contribution from other excitations, can
also substantially slow the condensate's decay, especially once
$\ncond$ has already gotten smaller.

We actually see the consequences of this decay process in
\Fig{power}.  In the upper left panel, the power spectrum has a bump
around $ak=0.9$ which moves to lower $k$ at late time as $\omega$
shrinks (since $\omega \propto \ncond^{1/3}$).  These are the
condensate decay products which have not yet rescattered into other
wave numbers -- as already realized in \cite{berg2}.  The same peak
does not occur for $N=2$, as the figure shows and for reasons we will
now see.

The situation is quite different for $N \geq 2$ because the condensate
is maximally charged.  The process in \Fig{42diagrams} cannot occur if
all 4 initial particles are from the condensate, because all incoming
particles have the same $O(N)$ charge, so the process
violates charge conservation and the relevant diagrams vanish
identically.  Instead, 3 condensate particles must
pick up one particle of opposite charge from the thermal bath to
produce two final-state particles with the same charge as the
condensate.  These are then free to propagate to a different region to
keep the fluctuations locally charge-neutral.  In \Eq{decayest1}, one
of the $\int d^3 p f_p$ must return the number of normal excitations,
$\ncond \to  \omega^3 f(\sim 1 \omega)$.  The decay rate of the
condensate is estimated as
\be
\frac{d\ncond}{dt} \sim \lambda^4 \frac{\ncond^3}{\omega^5}
f^2(\sim 1 \omega)
\ee
which is smaller than the previous estimate by a factor of
$\sim \omega^3 f(\sim 1 \omega)/\ncond$, which is the ratio of
infrared but non-condensate excitations to condensate excitations.
The condensate should decay much more slowly for
$N\geq 2$ than for $N=1$, especially in the regime where the
condensate is large.  It also does not decay into particles of a
specific $k$ value, since there is an energy and momentum range for
the particle picked up from the medium -- hence the absence of a peak
around $k\sim \sqrt{3}\omega$ in the upper-right plot in \Fig{power}.

\begin{figure}[ht]
  \putbox{0.45}{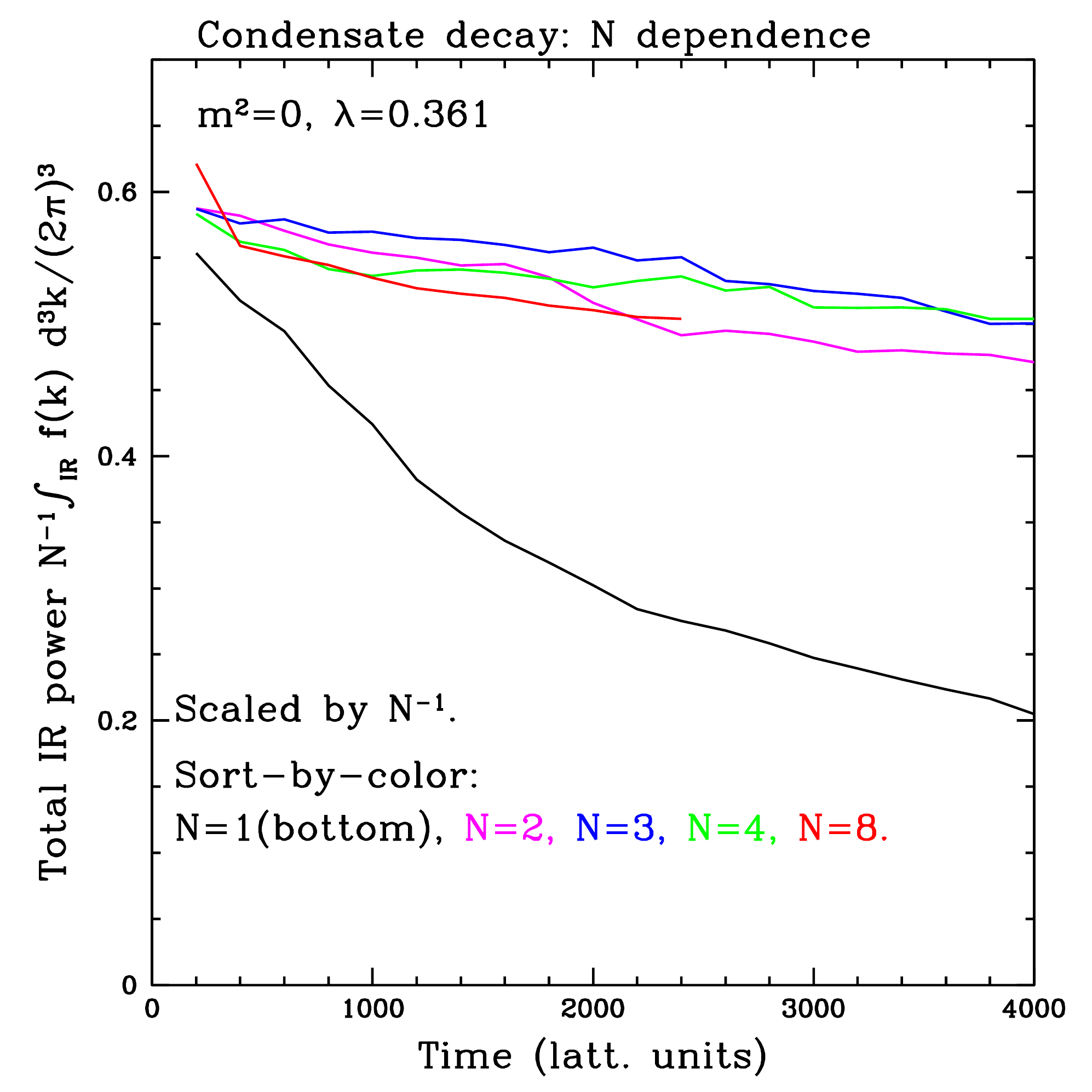} \hfill \putbox{0.45}{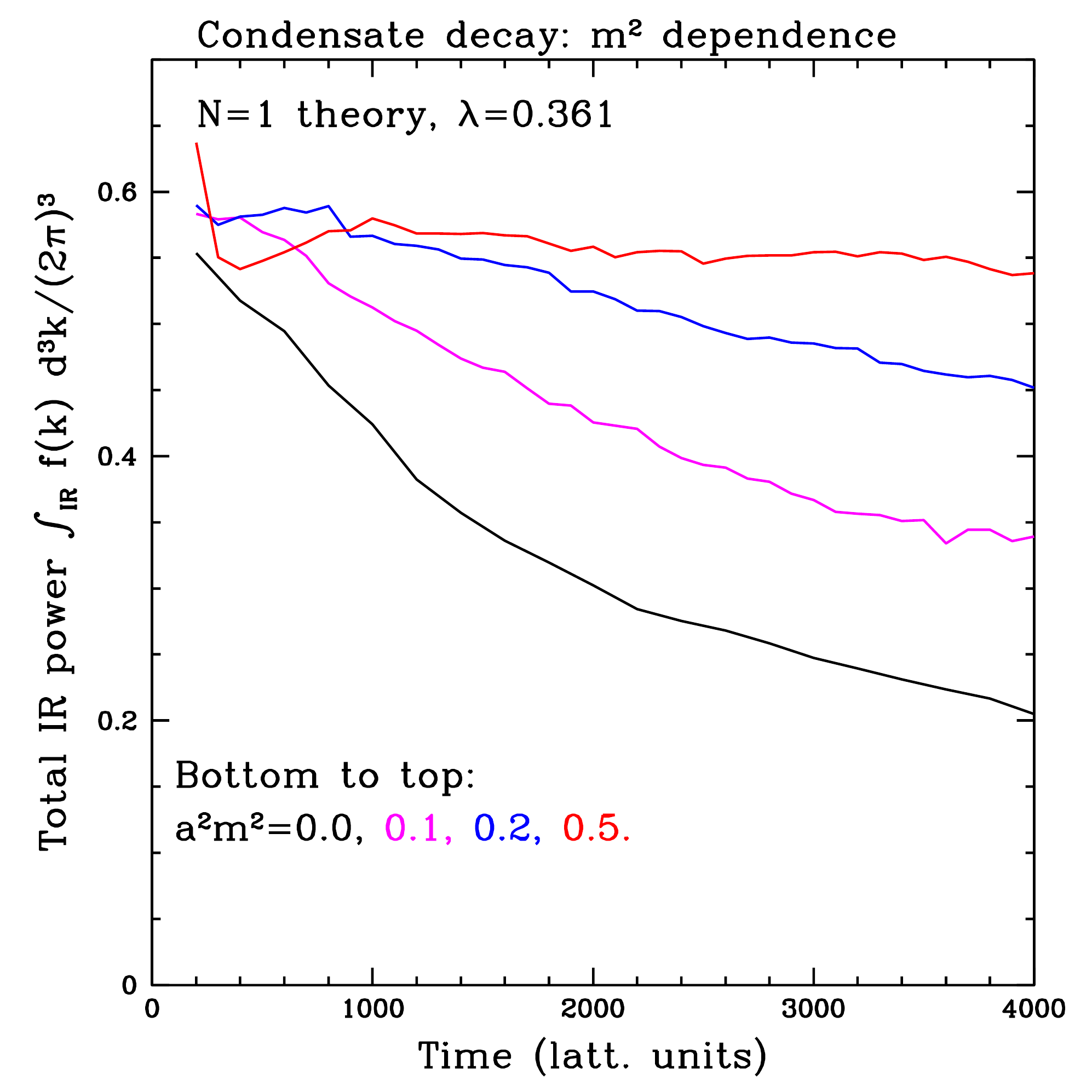}
  \caption{\label{cond_decay}
    Condensate as a function of time for the $N=1,2,3,4,8$ theories,
    left; condensate as a function of time as the explicit $m^2$
    term is increased for the case $N=1$, right.}
\end{figure}

We explore condensate decay as a function of $N$
in \Fig{cond_decay}, which shows the particle number
in the condensate (total particle number at small $k$ as
defined in Appendix \ref{lattdetail})
as a function of time for the $N=1,2,3,4,8$
theories.  The figure shows a stark difference between the $N=1$ case,
where there is no charge-conservation constraint, and the $N>1$ cases,
where there is.  Note that the initial particle number, and the
particle number which needs to move into the condensate, scales as $N$
for our initial conditions; we have explicitly removed this factor in
the figure.  We also study the effect of including an explicit $m^2$
term, which raises the oscillation frequency and therefore
suppresses the condensate decay according to \Eq{decayest1}, for the
$N=1$ case.  For the parameters studied, the $m^2=0$ case has
$a^2\omega^2 \sim 0.3$, so the first two $m^2>0$ curves still have the
oscillation frequency dominated by the condensate's self-interactions;
nevertheless, the addition of an explicit $m^2$ term raises $\omega^2$
enough to substantially suppress the condensate's decay.
We have not yet tried to explain these curves
quantitatively, though it would be interesting to do so in future.
They certainly support qualitatively the expectations explained
above.

\section{Discussion}
\label{discussion}

We have considered the deep IR behavior of classical scalar field
theory, where the approximate conservation of particle number causes
the development of a particle-number storing condensate.  Because
condensate formation occurs locally throughout the system, the
condensate is initially incoherent, and must develop long-range
order.  Our most interesting result is that, in the case that the
condensate is large enough that its self-interactions dominate its
oscillation frequency, the condensate's local structure is
qualitatively different in single-component ($N=1$) theory than in
$O(N)$ symmetric $N\geq 2$ component theories.  In the latter case,
the condensate locally carries the maximal possible charge density,
with overall charge neutrality maintained by spatial inhomogeneity of
the condensate.  This inhomogeneity persists indefinitely in finite
volume, and the charge density impedes the decay of the condensate,
which is much slower for $N>1$ than for $N=1$.  Finally, we have
considered the large-$k$ tail of the condensate, which appears to fall
as a power law which is \textsl{not} determined by the topological
structures present due to the long-range disorder of the condensate.

\section*{Acknowledgments}

I would like to thank Thomas Epelbaum and J\"urgen Berges for helpful
conversations, Johannes Walcher for help on homotopy for some specific
spaces, and Paul Mercure for his patience.  

\appendix

\section{Lattice details}
\label{lattdetail}

We give a few details of our lattice implementation.  We discretize
the theory of \Eq{Lagrangian} on the lattice with a standard nearest-neighbor
implementation of the gradient term and leapfrog update rule, with
temporal step equal to $1/10$ of the spacing.  Our initial conditions
are $\phi_a=0$ and each $k$-mode $\dot{\phi}_a(k)$ drawn from a
Gaussian distribution of fixed width for $k < \kinit$ (technically,
$\tilde{k}^2 \equiv \sum_{i=1,2,3} 2-2\cos(k_i a) < \kinit^2$) and zero for
$k>\kinit$.  In practice we use $\kinit = 1/a$, $a$ the lattice
spacing.  This provides large occupancy per unit energy, so that as the
energy fills into all available modes and in particular into
higher-$k$ modes, there is excess particle number which must fill the
IR condensate.  We choose $m^2=0$ except in a few simulations where we
study its effect on the condensate's time evolution.

The field normalization and $\lambda$ value overdefine the
system; one is free to rescale $\phi \to \xi \phi$, with $\xi$ a
constant, simultaneously rescaling $\lambda \to \xi^{-2} \lambda$.
We normalize the field so the total energy is $N$ per lattice site.
We chose this convention because, to the extent that the system is
weakly coupled, the eventual equilibrium (after the condensate decays)
has energy equipartitioned between kinetic and potential+gradient
energy, and this corresponds to an energy of $\half$ per degree of
freedom, and hence to a temperature of 1.  With this convention,
our simulations all used $\lambda=0.361$.
This value kept the thermal oscillation frequency somewhat below
$a\mth=1$ so that the 
IR behavior should not be too contaminated by lattice effects but the
dynamic range is still relatively good.

When we Fourier transform the $\phi$ and $\dot\phi$ fields to form a
power spectrum, we treat $k^2 < 8/aL$ separately from $k^2 > 8/aL$
(where $a$ is the lattice spacing and $L$ is the box length).  For the
former, we bin all Fourier components with the same value of $k^2$,
\textsl{eg}, $k=\frac{2\pi}{L}(2,2,1)$ is combined with its
cubic-invariance equivalents and with $k=\frac{2\pi}{L} (3,0,0)$ but
not with $(3,1,0)$.  For $k^2>8/aL$ we use $\tilde k^2$ bins
of width $2/aL$.

We define a range of wave numbers as being the range which contains
the condensate by finding the bin with the largest average Fourier
power in
$\omega^2 \langle \phi^2(k)\rangle + \langle \dot{\phi}^2(k)\rangle$,
and considering all bins where this quantity is at least $1/300$ of
this maximum value.  We then find the integral of
$\langle \phi^2(k) \rangle$ and $\langle \dot\phi^2 \rangle$ over this
range, and find $\omega^2$ self-consistently as
\be
\label{omsqdef}
\omega^2 \equiv \frac{\sum_{\mathrm{cond.modes}} \dot\phi^2(k)}
      {\sum_{\mathrm{cond.modes}} \phi^2(k)} \,,
\ee
re-determining the range of modes in the condensate until we reach
self-consistency.  The occupancy at $k$ is estimated as
\be
\label{fdef}
f(k) = \omega \langle \phi^2(k) \rangle
+ \omega^{-1} \langle \dot\phi^2(k) \rangle \,.
\ee
These steps assume that the condensate is carried in a range of $k$
such that $k^2 \ll \omega^2$, which becomes true quite quickly; if it
is not true it does not make sense to speak of a local condensate.
Finally, we define the characteristic wave-vector for the local
condensate as
\be
\label{kconddef}
k_{\mathrm{cond}} = \frac{\sum_{\mathrm{cond.modes}} k f(k)}
{\sum_{\mathrm{cond.modes}} f(k)} \,.
\ee

\section{Oscillation frequencies}
\label{appomega}

Here we complete the details in the calculation of oscillation
frequencies for single-component and ``circular'' multi-component
oscillation.

Consider a scalar oscillating back-and-forth in a $(\lambda/8) \phi^4$
potential.  If the field oscillates with amplitude $\phi_0$ then the
energy is
\be
\label{N1en}
\varepsilon = 
\frac{\lambda}{8} \phi_0^4 = \frac 12 \dot\phi^2 + \frac{\lambda}{8}
\phi^4 \qquad \rightarrow \qquad
\dot\phi = \frac{\sqrt\lambda}{2} \sqrt{ \phi_0^4 - \phi^4 }
\ee
and the period of the oscillation is
\bea
\label{N1T}
\frac{T}{4} & = & \int_0^{\phi_0} \frac{dt}{d\phi} d\phi
 = \int_0^{\phi_0} \frac{1}{\sqrt{2\varepsilon-2V(\phi)}} d\phi
 = \frac{2}{\phi_0 \sqrt{\lambda}}
  \int_0^1 \frac{dx}{\sqrt{1-x^4}}
\nonumber \\
 & = & \frac{\sqrt{\pi} \Gamma(1/4)}{2 \Gamma(3/4) \phi_0 \sqrt{\lambda}}
\eea
which gives a frequency of%
\footnote{%
  This is not the same as the naive estimate
  $\omega^2 = \frac{3\lambda}{2} \langle \phi^2 \rangle$,
  with
  $$
  \langle \phi^2 \rangle = \frac{\int_0^{\phi_0} d\phi \frac{\phi^2}
    {d\phi/dt}}{\int_0^{\phi_0} d\phi \frac{1}{d\phi/dt}} =
  \frac{\phi_0^2 \int_0^1 dx\,x^2/\sqrt{1-x^4}}
          {\int_0^1 dx/\sqrt{1-x^4}}
          = \phi_0^2 \frac{\Gamma^2(3/4)}{\Gamma(5/4)\Gamma(1/4)}
          $$
  which gives $\omega = .8279 \sqrt{\lambda} \phi_0$.  This estimate
  may work for larger-$k$ excitations, but it does not work at $k=0$
  because the field is time-coherent with itself.
  }
\be
\label{N1omega}
\omega = \frac{2\pi}{T} = \frac{\sqrt{\pi} \Gamma(3/4)}{\Gamma(1/4)}
 \sqrt{\lambda} \phi_0 \simeq 0.59907 \sqrt{\lambda} \phi_0 \,.
\ee

We can re-cast this in terms of the energy density using
$\varepsilon = \lambda \phi_0^4/8$:
\be
\label{N1omega2}
\omega = \frac{(8\pi^2)^{1/4} \Gamma(3/4)}{\Gamma(1/4)}
\lambda^{1/4} \varepsilon^{1/4}
= 1.00751 \lambda^{1/4} \varepsilon^{1/4} \,.
\ee

When we add one particle to the condensate,
we add an energy of $\omega$.
So the particle number stored in the condensate is%
\footnote{%
  If this looks strange, recall the behavior of a quantum system with
  potential $V$ in the WKB approximation, valid for high levels.
  The level number $f$ is given in terms of the level energy $E$ by
  $f = \frac{1}{\pi}\int_{\phi|V(\phi)<E} \sqrt{2(E-V)} d\phi$.  The
  level spacing $dE/df$ sets the oscillation frequency of a
  superposition of levels,
  So the period is given by $T = 2\pi/\omega = 2\pi df/dE$
  which is $T = 4\int (2(E-V))^{-1} d\phi$, agreeing with \Eq{N1T}.}
\be
\label{N1neps}
n(\varepsilon) = \int_0^{\varepsilon} \frac{d\varepsilon'}{\omega}
= \frac{4}{3} \frac{\varepsilon}{\omega} = 
\frac{4}{3 (1.00751) \lambda^{1/4}} \varepsilon^{3/4} \,.
\ee
Alternatively,
\be
\label{N1epsn}
\varepsilon \simeq 0.68825 \lambda^{1/3} n^{4/3} \,.
\ee

Now consider instead a scalar in 2 or more components, with $\dot\phi$
orthogonal to $\phi$ and large enough that the field's amplitude
remains fixed at $\phi_0$ while changing direction in field space.
This is like circular orbital motion in an $r^4$ potential.  The energy
density is 
\be
\label{n2eps1}
\varepsilon = \frac 12 \dot{\phi}_0^2 + \frac{\lambda}{8} \phi_0^4
\ee
and the Virial relation
$\frac{1}{2} \dot{\phi}^2 = 2 \frac{\lambda}{8} \phi_0^4$ gives
\be
\label{n2eps2}
\varepsilon = \frac{3}{4} \dot\phi_0^2 = \frac{3\omega^2}{4} \phi_0^2
 = \frac{3\lambda}{8} \phi_0^4
 \qquad \mbox{and} \quad
 \omega^2 = \frac{\lambda}{2} \phi_0^2 
 \ee
from which we easily find
\be
\label{n2omega}
\omega = \frac{2^{1/4}}{3^{1/4}} \lambda^{1/4} \varepsilon^{1/4}
 \simeq 0.9036 \lambda^{1/4} \varepsilon^{1/4} \,.
\ee
Following the same steps as before, we find
\be
\label{n2neps}
n(\varepsilon) = \frac{4 \varepsilon}{3\omega} = 
\frac{2^{7/4}}{3^{3/4} \lambda^{1/4}} \varepsilon^{3/4}
\ee
which equals $n=\phi_0 \dot{\phi}_0 = \omega \phi_0^2$ as expected,
and
\be
\label{n2epsn}
\varepsilon \simeq \frac{3}{2^{7/3}} \lambda^{1/3} n^{4/3}
 \simeq .59528 \lambda^{1/3} n^{4/3} \,.
\ee
Therefore the energy associated with a condensate which makes a circular
rotation in field space is lower, at fixed number density, than the
energy for the condensate to oscillate straight back-and-forth.

\bibliographystyle{JHEP}
\bibliography{cond}

\end{document}